\documentclass[nofootinbib,11pt]{revtex4-1}
\usepackage{amsmath,amssymb,pgf,pgfarrows,pgfnodes,float,appendix, hyperref}
\usepackage{graphicx}
\usepackage{subfigure}
\usepackage[margin=0.9in]{geometry}

\newcommand{\bea}{\begin{eqnarray}}
\newcommand{\eea}{\end{eqnarray}}

\begin{document}
\title{Hilbert Bundles and Holographic Space-time Models}
\author{T. Banks}
\affiliation{NHETC and Dept. of Physics and Astronomy, Rutgers University, Piscataway, NJ }

\begin{abstract} We reformulate Holographic Space-time Models in terms of Hilbert bundles over the space of time-like geodesics in a Lorentzian manifold.  This reformulation resolves the issue of the action of non-compact isometry groups on finite dimensional Hilbert spaces. Following Jacobson\cite{ted1995} we view the background geometry as a hydrodynamic flow, whose connection to an underlying quantum system follows from the Bekenstein-Hawking relation between area and entropy, generalized to arbitrary causal diamonds.  Time-like geodesics are equivalent to nested sequences of causal diamonds and the area of the holoscreen\footnote{The holoscreen is the maximal $d-2$ volume ("area") leaf of a null foliation of the diamond boundary. We use the term area to refer to its volume.}
encodes the entropy of a certain density matrix on a finite dimensional Hilbert space.
We review arguments\cite{carlip}\cite{solo}\cite{bz} that the modular Hamiltonian of a diamond is a cutoff version of the Virasoro generator $L_0$ of a $1 + 1$ dimensional CFT of large central charge, living on an interval in longitudinal coordinate on the diamond boundary.  The cutoff is chosen so that the von Neumann entropy is ${\rm ln D_{\diamond}} ,$ up to subleading corrections, in the limit of large dimension diamond Hilbert space.   We also connect those arguments to the derivation\cite{V291} of the 't Hooft commutation relations for horizon fluctuations. We present a tentative connection between the 't Hooft relations and $U(1)$ currents in the CFT's on the past and future diamond boundaries.  The 't Hooft relations are related to the Schwinger term in the commutator of vector and axial currents.  The paper\cite{V291} can be read as a proof that near horizon dynamics for causal diamonds much larger than Planck scale is equivalent to a topological field theory of the 't Hooft CR, plus small fluctuations of transverse geometry.  Connes'\cite{connes} demonstration that Riemannian geometry is encoded in the Dirac operator leads one to a completely finite theory of transverse geometry fluctuations, in which the variables are fermionic generators of a superalgebra, which are the expansion coefficients of sections of the spinor bundle in Dirac eigenfunctions.  A finite cutoff on the Dirac spectrum gives rise to the area law for entropy, and makes the geometry both "fuzzy"\cite{tbjk} and quantum.  Following the analysis of Carlip and Solodukhin, we model the expansion coefficients as two dimensional fermionic fields.  We argue that local excitations in the interior of a diamond are constrained states where the spinor variables vanish in regions of small area on the holoscreen. This leads to an argument that quantum gravity in asymptotically flat space must be exactly supersymmetric.

\end{abstract}

\maketitle
\tableofcontents

\section{Introduction}

Quantum Field Theory (QFT) is our most successful model of the physics of the universe and enjoys a remarkable degree of agreement with experiment.  Every experiment ever done, or likely to be done in the forseeable future, is performed by a detector moving for a finite amount of proper time along a near geodesic time-like trajectory in a Lorentzian manifold with curvature that is small in Planck units.  Such an experiment defines a causal diamond. The authors of\cite{CKN} showed that the agreement of QFT with current experiments was unaffected if we omitted from the QFT Hilbert space all states whose semi-classical gravitational back reaction would create a black hole larger than the causal diamond.  With the UV/IR cutoffs imposed in\cite{CKN} current experiments were near the limits implied by these constraints, but\cite{pdetal} showed that other forms of cut off put experimental tests of these {\it a priori} constraints on QFT off into the distant future.  The total entropy allowed by these constraints is $\sim S_{BH}^{\frac{d-1}{d}}$, in $d$ dimensional space-time.  $S_{BH}$ is the area ($d-2$ volume) of the maximal area $d-2$ surface on the null boundary of the diamond, divided by $4G_N$.   We will call this the area of the holographic screen.

Although the experimental success of QFT does not require us to believe this, the theoretical structure of QFT gives us a compelling reason to suspect that there is something special about an area law for the entropy of a causal diamond.  In the 1960s, algebraic quantum field theorists\cite{haagbrandeis} discovered that the algebra of operators localized in a diamond was always Type III in the Murray von Neumann classification.  This means that the algebra has no density matrices of finite entropy.  In 1983, Sorkin\cite{sorkin} computed the vacuum entanglement entropy of a diamond and found that it was proportional to the area of the holographic screen, with a UV divergent proportionality constant.   This calculation languished in obscurity until it was repeated by more sophisticated methods in the 1990s\cite{srednicki}\cite{cw}. It is clear that these calculations are unchanged in any state obtained by acting with a finite number of smeared local operators on the vacuum, as long as their support is spacelike separated from the boundary of the diamond.  The infinite entanglement comes from divergent light cone commutators of fields on either side of the diamond boundary.  

Entanglement entropy of one system with another is a lower bound on the maximal entropy of the system.  The fact that the finite temperature entropy of the portion of a Cauchy slice whose causal boundary is the diamond, is finite after vacuum entropy subtraction, combined with the Cohen Kaplan Nelson (CKN) argument, is a strong hint that this divergent boundary entropy has fundamental significance. Susskind and Uglum\cite{sussug} and Jacobson\cite{jacobson} suggested that this divergence should be viewed as a renormalization of Newton's constant in the Bekenstein Hawking entropy law.  None of these authors commented on the revolutionary nature of this suggestion.  The Bekenstein-Hawking law was a proposal about black holes.  The calculations of\cite{sorkin}\cite{srednicki}\cite{cw} were all done in empty flat space.  At the time, the present author found this confusing, but was not clever enough to follow up on the hint.  
 
 The most convincing evidence that we should take the area law proposal for the entropy of a causal diamond seriously came one year later in the work of Jacobson\cite{ted1995}.  This paper was phrased in terms of local changes of entropy and so did not explicitly invoke the {\it Covariant Entropy Principle}(CEP): {\bf In models of quantum gravity, the entropy of a causal diamond is equal to the area of its holographic screen divided by $4G_N$. }  Entropy should be thought of as the expectation value of the modular Hamiltonian associated with the "empty diamond" state.  The CEP is meant as the first term in an asymptotic expansion of the entropy as a function of the area.  The term "area" makes sense only in the context of this expansion.  It was not until 1998 that the CEP was formulated explicitly, first in the special case of FRW cosmology by Fischler and Susskind\cite{fs} and then in the masterful follow up work by Bousso\cite{bousso}.
 
 Implicit in the CEP is the central notion proven in\cite{ted1995}:  the equations
 \begin{equation} k^m (x) k^n (x) (R_{mn} (x) - \frac{1}{2} g_{mn}(x) R(x) - 8\pi G_N T_{mn}(x)) = 0 , \end{equation} are the hydrodynamic equations of the CEP.  Here $k^m (x)$ is an arbitrary null vector in space-time. That is, these equations follow from the definition of Lorentzian geometry (which includes the Raychaudhuri equation for the local spreading of a congruence of geodesics) and the covariant conservation of the stress tensor, applied to the thermodynamic equation $dE = T dS$, assuming that entropy is given by the CEP.  One also uses the local relation $E = k^m k^n T_{mn}$ and chooses a congruence of trajectories centered around a maximally accelerated Unruh trajectory, which grazes a point on the holographic screen of the diamond.  The fact that the CEP is only an asymptotic relation follows from the fact that the derivation assumes that everything can be treated as locally flat.  
 
 A few comments are in order.  The CEP can be "derived" from Euclidean path integrals\cite{BDF} just like black hole or dS entropy\cite{GH}.  The relation between Euclidean path integrals and hydrodynamics has been clarified recently in\cite{tbwormhole}.  For a system consisting of large independent subsystems $X_i$ connected by small "interfaces" with many fewer degrees of freedom, hydrodynamics can be derived from quantum mechanics as a Markov equation for the diagonal matrix elements of the density matrix in a basis of states in which the local values $C(X_i)$ of conserved quantities are simultaneously diagonal\cite{tbaletal}.  Kac's path integral solution\cite{Kac} of this Markov equation is the Euclidean path integral for hydrodynamics.
 
 The hydrodynamic view of Einstein's equations means that they are valid in contexts where there is no systematic weak coupling or large N expansion.  Strongly coupled condensed matter systems have high entropy states whose coarse grained properties are described by the Navier-Stokes equations, even though the underlying quantum mechanics is not well approximated by perturbative quasi-particle physics.  As a bonus, Jacobson's derivation makes it clear that the cosmological constant in Einstein's equations should not be thought of as an energy density since it does not appear in the equations of hydrodynamics.  String theorists should have known this since the invention of the AdS/CFT correspondence, where the c.c. plays the role of a parameter in the high energy density of states in a CFT and is determined by discrete parameters like the gauge group $SU(N)$.  
 
 Further evidence for the CEP, and for Jacobson's view of GR as hydrodynamics, comes from remarkable work done by Carlip\cite{carlip} and Solodukhin\cite{solo} in 1998\footnote{Carlip had begun exploring these ideas earlier\cite{carlipearly} but the fully developed realization of his ideas appeared simultaneously with the independent paper of Solodukhin.}.  These authors proposed a state counting interpretation of black hole entropy for quite general black holes.  More recently, K. Zurek and the present author realized that their derivation applied to a general causal diamond\cite{bz}.  We will explore these ideas using an argument introduced by the Verlindes\cite{V291} in order to provide a formal derivation of 't Hooft's commutation relations between light ray operators in high energy scattering.  The same argument is precisely suited for studying the near horizon limit of any causal diamond.  In a later section, we will return to provide a microscopic model of the 't Hooft commutation relations.
 
 \subsection{$Verlinde^2 \rightarrow Carlip + Solodukhin$ in a General Causal Diamond}
 
 The observation of the Verlindes is that one can analyze high energy scattering at large transverse distance by taking length scales in a two dimensional subspace of Minkowski signature to be Planck scale, while those in the transverse dimensions are of order $L \gg L_P$.  The same is true in the near horizon limit of any causal diamond, since the two dimensional length scales are nearly null and the transverse length scale $L$, will generally be much larger than $L_P$.  The Einstein Hilbert action splits into three terms of very different size.
 \begin{equation} I_{EH} = (L/L_P)^{d-4} I_{tH} + (L/L_P)^{d - 2} I_{\perp} + I_3. \end{equation}
 \begin{equation} I_{\perp} = \int \sqrt{-g}\sqrt{h} [-R(g) - g^{ab}\partial_a h_{mn} \partial_b h_{kl}(h^{ml}h^{nk} - h^{hm}h^{nl})] . \end{equation}
  \begin{equation} I_{tH} = \int \sqrt{-g}\sqrt{h} [-R(h) - h^{mn}\partial_m g_{ab} \partial_n g_{cd}(g^{ad}g^{bc} - g^{ac}g^{bd})] . \end{equation}
  We will not need the formula for the third term.
  Here $g_{ab} = \hat{g}_{ab} - \hat{g}_{ai}h^{ij}\hat{g}_{jb}$, where $\hat{g}$ is the original metric. 
  The equations of motion of $I_{\perp}$ say that $g_{ab}$ is flat, and that $h_{mn}$ is independent of $y$.  The authors of\cite{V291} then show that $I_{tH}$ becomes a purely topological action for the boundary values of the coordinate transformation that takes $g_{ab}$ to $\eta_{ab}$.  Those boundary values satisfy the 't Hooft commutation relations.  We will return to a microscopic model of those commutation relations below. 
  
  A quick and dirty way to understand the argument of\cite{V291} and to see that it is valid in any dimension, is to note that we can view the near horizon limit of any causal diamond as an asymmetric rescaling of coordinates in which the transverse coordinates are taken much larger than those in the two dimensional Minkowksi signature directions.  Under this rescaling, the covariant components of the Ricci tensor transform as a tensor and its clear that the dominant term for large $(L/L_P)$ is $R_{ab}$ .  The leading terms in Einstein's equations, for a stress tensor that consists of pure traceless boundary waves (as suggested by the CEP) are precisely those following from $I_{\perp}$.   
  
  The fact that the fluctuations of the transverse metric become small whenever length scales are large compared to the Planck scale is another indication that Jacobson's hydrodynamic view of gravity is correct.  One can construct a very general derivation of hydrodynamics for lattice quantum systems, using only the assumption that large blocks of the lattice have Hamiltonians that commute with each other up to subleading surface terms\cite{tbaletal}.  The Schrodinger equation leads directly to classical stochastic equations for the hydrodynamic variables, which have a standard Euclidean path integral solution.  Fluctuations of transverse geometry are similarly suppressed simply by the size of the region they describe.  No assumptions of "weak string coupling" are necessary.
  
  Solodukhin's analysis of the linearized fluctuations around the classical solution is particularly transparent.  One defines a scalar field $\phi$ in the Lorentzian dimensions in terms of the fluctuations of the conformal factor of the metric $h$ around its classical value.  After a $\phi$ dependent Weyl transformation of the two dimensional metric (which, to linear order in $\phi$ is just a constant rescaling) the action takes the form
  \begin{equation} I[\phi] = - \int d^2 y \sqrt{-g} [\frac{1}{2}g^{ab}\partial_a \phi \partial_b \phi + q \phi \sqrt{\frac{(d-3)S}{(d-2)8\pi}} R(g) + U(\phi)] .\end{equation} The metric is set to be Minkowski by the equations of motion of $I_{\perp}$, but we have kept it general to show that the field $\phi$ has a stress tensor with a large classical central charge.  The first two terms define a classical conformal field theory, which was shown in\cite{BPZ} to have a stress tensor whose correlation functions satisfy all the Ward identities of a general CFT with the same value of the central charge, because stress tensor correlators in two dimensional CFT are completely determined by the central charge.  This is often called the Liouville theory, though it does not have the usual two dimensional cosmological constant term that gives Liouville's equation for constant curvature metrics. An illuminating way to state the result of\cite{BPZ} is that the "Liouville" theory is the solution of the hydrodynamic equations of two dimensional CFT with a given value of $c$.  Solodukhin shows that the potential term $U(\phi)$ contains only "classically irrelevant" perturbations of this CFT, in the near horizon limit.  
  
  In\cite{AS}, Strominger argued that the boundary "Liouville" theory discovered by Brown and Henneaux\cite{brownhenneaux} in $2 + 1$ dimensional AdS gravity, was an avatar of the AdS/CFT correspondence.  In our Jacobsonian language: asymptotic 3d gravity in AdS space is the hydrodynamics of $1+1$ CFT.  Strominger argued, as a consequence, that quantum gravity should be quantum CFT.  This was an after the fact justification for the $AdS_3/CFT_2$ correspondence. Carlip and Solodukhin advocate using the same logic for all black holes and show that Cardy's entropy formula reproduces the Bekenstein-Hawking formula in every case.  In\cite{bz} we argued that these arguments applied to every causal diamond, giving further justification for the CEP.  In addition we argued that they implied the universal fluctuation formula $(\Delta K)^2 = \langle K \rangle$ for the modular Hamiltonian\footnote{The sole known exceptions are the horizons of large stable black holes in AdS space. These are also the only horizons that propagate sound modes.  Both facts can be understood in terms of the tensor network model of AdS/CFT.  The universal fluctuation formula is valid in nodes of the tensor network but the entropy of a large black hole is dominated by sound modes propagating between the nodes, which have a different capacity of entanglement.  See the Appendix for more detail.}.  Solodukhin remarks that there is no apparent reason for neglecting fluctuations of the unimodular part of the transverse metric $h_{mn}$.  In the next section we will use insights from the Holographic Space-time formalism to incorporate these into the Carlip-Solodukhin picture.
  
When combined with the CKN argument, the work of Carlip and Solodukhin leads one to a remarkable conclusion. Most of the quantum degrees of freedom of a model of quantum gravity are inaccessible to local experiments done on near geodesic trajectories through the diamond.  They reside on the boundaries of causal diamonds and we will only be able to get very indirect experimental information about them.   What is more, so far the CS picture does not give us a clue about the nature of the local excitations, to which the usual rules of QFT apply.  These are the excitations explored by experiments done along geodesics inside a causal diamond.  Fortunately, the answer to that question was found by Fiol, Fischler and the present author some time ago, and we will return to it below.

\section{Transverse Geometry in Terms of Fluctuating Fuzzy Spinors}
 
 The Holographic Space-time (HST) formalism was introduced in order to construct models of quantum gravity with a closer connection to local physics than that achieved in perturbative string theory or the AdS/CFT correspondence.  Its most basic postulate is the existence of a net of local subalgebras of the operator algebra describing an entire space-time.  These subalgebras are finite dimensional and are in one to one correspondence with finite area causal diamonds in the space-time.  When the area is much larger than any microscopic scale the dimension of the Hilbert space associated with a diamond is approximately\footnote{A more precise statement is that the area is the expectation value of the modular Hamiltonian of the empty diamond state in the Hilbert space. However, in models constructed so far, this coincides with the log of the dimension up to subleading corrections.}
\begin{equation} d ({\cal H}_{\diamond}) \approx e^{\frac{A}{4G_N}} . \end{equation}  A nested sequence of intervals along any time-like trajectory defines a nested sequence of causal diamonds.  Every diamond contains a unique timelike geodesic that connects its past and future tips. Given a choice of nested intervals on a timelike geodesic we have a sequence of causal diamonds.  The relations between proper time interval and area for all diamonds along all geodesics completely determines the space-time metric $g_{mn}$.  The relation between area and Hilbert space dimension thus allows us to construct a map from quantum mechanics to geometry:  we consider a {\it Hilbert bundle} over the space of time-like geodesics that connect two Cauchy slices in the space-time\footnote{For space-times that are asymptotically AdS, we must restrict the maximal proper time between points on the two Cauchy surfaces, so that the maximal causal diamonds have finite area, in order to treat the infinite area limit with proper care.  Note that for non-negative c.c. we either have no restriction on the proper time between past and future surfaces or only the restriction that it be finite. }.  We can make a synchronized choice of proper time intervals,starting from the past Cauchy surface or going backwards from the future surface.  In time symmetric space-times we can nest outwards from a small diamond surrounding a point of time symmetry (Figure 1). 
 \begin{figure}[h]
\begin{center}
\includegraphics[width=01\linewidth]{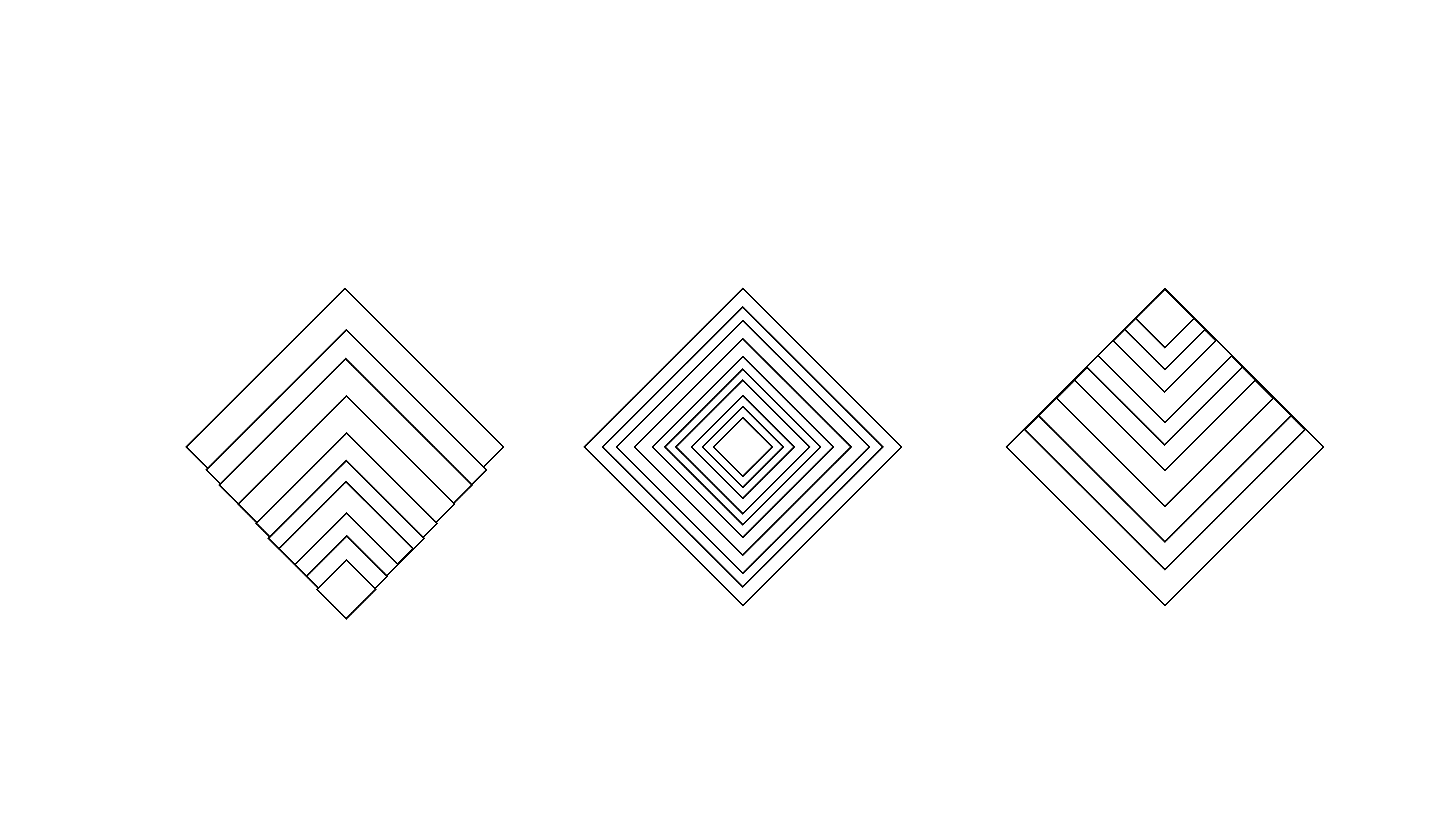}

\caption{Future directed, time symmetric, and past directed nested coverings of a causal diamond.} 
\label{fig:nestedcoversofadiamond}
\end{center}
\end{figure}

Specifying the dimensions of Hilbert spaces for each time interval is equivalent to specifying the space-time metric, in the limit of exponentially large dimensions.  The basic hypothesis of HST is that Hilbert space dimension is the correct extension of the concept of area of a causal diamond's holographic screen, to short distances.  This idea lines up perfectly with Jacobson's\cite{ted1995} claim that Einstein's Equations are the hydrodynamic equations of the area law, and provides an immediate resolution of space-like singularities in solutions of GR.  Space-like singularities\cite{Hawking}\cite{penrose} are places where causal diamonds shrink to zero area and it's precisely in low entropy situations that we can expect the hydrodynamic approximation to a quantum system to break down.  

The hydrodynamic view of Einstein's equations casts doubt on a number of assumptions common among researchers in all branches of quantum gravity.  The first of these is that there should be a single "background independent" formulation of quantum gravity, with all particular instances of the theory some sort of "vacuum states".   Among string theorists, this myth should have been dispelled long ago by the success of the AdS/CFT correspondence.  From the hydrodynamic perspective it is tantamount to saying that because almost all systems have states that satisfy the Navier-Stokes equation, then they are described by the same microscopic quantum Hamiltonian.  

The second erroneous idea, most common among string theorists, is that the Einstein equations should always be viewed as the first term in some kind of semi-classical expansion, which is in some way related to a "path integral over metrics".
For condensed matter systems with a gapless ground state, it is often the case that one can view the low lying excitations as quantized fields obeying the linearized hydrodynamic equations, with interactions that can be understood by Wilsonian renormalization group analysis and symmetry considerations.   The classical hydrodynamic equations have a much broader validity than this and apply to states far removed from the ground state, in which it is inappropriate to quantize them.  Instead, one searches for a microscopic system that reproduces the same coarse grained hydrodynamic flow.   This is the principle that we will be following.  Our models of quantum gravity always begin with a fixed classical background, which we view as a hydrodynamic flow to which we must match a quantum theory, following the clues outlined in the previous section.  Hydrodynamic equations have to be supplemented by stochastic "stirring" terms, in order to agree with observation.  There is a long history of representing the solutions of these stochastic equations by Euclidean path integrals, following Kac' original treatment of the diffusion equation.  This is the way one should view Euclidean quantum gravity.

As noted, the big lacuna in the discussion of the previous section was a model for the fluctuations in the transverse geometry.  Here we follow the analysis of HST.  Finite entropy implies that the algebra of operators in a diamond be Type II in the Murray-von Neumann classification, but it is essentially impossible for experimental physics to distinguish between infinite Type II algebras and finite dimensional algebras, so HST has always been restricted to finite dimensional Hilbert spaces.  Any such Hilbert space is a representation of a fermionically generated superalgebra, or equivalently of a system of canonical fermionic oscillators with constraints.  Alain Connes pointed out long ago\cite{connes} that all of Riemannian geometry was encoded in the Dirac operator.  Indeed, on page 544 of\cite{connes} one can find a (rather abstract) formula for the geodesic distance between two points in a manifold in terms of properties of the Dirac operator.  Physicists are more familiar with the fact that the short time expansion of the heat kernel of the square of the Dirac operator can be written in terms of curvature invariants, and that  topological properties of the manifold are also related to the index of the Dirac operator.
This means that a cut off on the spectrum of the Dirac operator provides a method of "fuzzifying" a Riemannian geometry even if the space does not have a symplectic structure\cite{tbjk}. Furthermore, if we postulate that the expansion coefficients of a generic section of the spinor bundle into Dirac eigenfunctions of a {\it fixed} background metric, are $1 + 1$ dimensional (cut-off) quantum fields, then each quantum field theory state will define a probability distribution for the curvature invariants of the transverse geometry. 
More precisely, the quantum field 
\begin{equation} \Psi (x,y) = \sum_E e^{- i Et} \psi_E (y) \chi_E (x) , \end{equation} 
satisfies the Dirac equation, with Hamiltonian 
$H = [L_0, ] + D$ , where $D$ is the transverse Dirac operator in the background geometry, and $E$ its cutoff spectrum.  $[L_0, ]$ is the action of the $L_0$ generator on operators in the CFT.  Its spectrum is determined by that of $L_0$. Thus, we can view the fluctuations in $L_0$ as being fluctuations in the transverse Dirac spectrum, and thus of the transverse geometry. 

This is not the only possible definition of a fluctuating transverse geometry.   The fermion bilinears 
\begin{equation} \bar{\psi} (\sigma , x) \Gamma_{(A_{1}\ldots A_{k})} e_{m(1)}^{A_1} \ldots e_{m(k)}^{A_k} \psi (\sigma , x) , \end{equation} are rank $k$ differential forms, and the bundle of all these forms satisfies the Kahler-Dirac equation
\begin{equation} (d - d^{*}) F = 0 , \end{equation} where $d$ is the exterior derivative on the transverse manifold and $d*$ its Hodge dual.   This equation implies that there is a linear combination of the bilinear
\begin{equation} \bar{\psi} (\sigma , x) \Gamma_{(A_{1}\ldots A_{4})} e_{m(1)}^{A_1} e_{m(2)}^{A_2} \psi (\sigma , x) , \end{equation} and the divergence of the rank $5$ bilinear, which has the symmetry properties and satisfies the Bianchi identity of the field strength of an $O(d-2)$ connection on the spinor bundle.  This is an operator quantum field and we can view it as a fluctuating curvature tensor for the holoscreen.   It's not clear which, if any of these definitions is correct, or whether there is any way of testing either hypothesis.   We've emphasized that the kinds of measurements available to a local observer use detectors with far fewer q-bits than the number that actually describe the holoscreen.  We should not be surprised to discover that our experiments will never be able to probe the details of "quantum geometry".  

  If we arrange that the quantum density matrix for each causal diamond satisfies
\begin{equation} \langle K \rangle = \frac{A}{4G_N} , \end{equation}
\begin{equation} \langle (K - \langle K \rangle)^2 \rangle = \frac{A}{4G_N}, \end{equation} then we will have a quantum model whose hydrodynamic equations coincide with the solution of the Einstein equations from which we began, according to the analysis of Carlip and Solodukhin.

Another immediate benefit of adopting the Connes-Dirac formulation of Riemannian geometry and the quantization rules of HST is that it builds in the spin statistics connection.  The connection between spin and statistics is a theorem in local quantum field theory, but even more importantly, it is a property of the real world, and {\it local quantum field theory is not}.
We'll see below that in asymptotically flat space-time it also (almost) proves the necessity for Supersymmetry (SUSY) in asymptotically flat space: a "phenomenological" fact about string theory with no obvious explanation.  

We begin to see the outlines of a quantum theory of gravity, applicable on scales much larger than the Planck scale.  One starts from a classical space-time, obeying Einstein's equations with a stress tensor satisfying the null energy condition\footnote{The null energy condition follows from the second law of thermodynamics in Jacobson's derivation of the Einstein equations.}.  For each time-like geodesic in the space-time one introduces a nested sequence of proper time intervals and constructs a corresponding Hilbert bundle of cutoff $1 + 1$ dimensional fermion fields.  The UV and IR cutoffs in two dimensions are related to the minimal size of causal diamond (in Planck units) for which one believes the Carlip/Solodukhin arguments are valid.  At the moment, we have no principled way of deciding how large the minimal diamond is.  This choice of course removes any Big Bang singularities from the background space-time.  Once this choice is made, one increases the number of fermion fields according to the Bekenstein-Hawking-Carlip-Solodukhin area/entropy law.   The quantum state of an empty diamond has modular Hamiltonian equal to the $L_0$ generator of a $1+1$ CFT constructed from the fermions, with the UV/IR cutoffs described above.    The cutoff is imposed on the spectrum of $L_0$, with a central value chosen by the solution of the CS Liouville equation.  In general (see below) there appears to be a conformal manifold of CFTs to choose from and it is not clear whether these correspond to different consistent models for quantum fluctuations around the same classical geometry.  

\subsection{Time Evolution}

For causal diamonds in maximally symmetric spaces, the $L_0$ generator used by 
Carlip and Solodukhin showed up in more recent work of Casini, Huerta and Myers\cite{CHM} and Jacobson and Visser\cite{JV} on the modular Hamiltonian of conformal field theory on these space-times.  $L_0$ is a quantum operator, which implements the action of a certain space-time vector field on the time-slice through the holographic screen of the diamond.  The vector field preserves the diamond and the flows associated with it are time-like inside it and define a set of inextensible Diamond Universe (DU) coordinates inside the diamond.  In fact this is a conformal Killing vector of the maximally symmetric space-time, so the same statements will be true for any geometry conformal to the maximally symmetric space-times.  The DU coordinates will be different for different conformal factors. The Virasoro algebra fixes the normalization of the CKV, which is of course ambiguous as a geometrical object.

Now let's consider two causal diamonds in a future directed nesting, with future tips separated by a single Planck unit.  We have argued that each should be described by a $1+1$ dimensional CFT built from fermion fields, with the larger diamond containing more fermion fields than the smaller one.  The density matrix for the empty diamond state
is $e^{-L_0 (\tau)}$ for the small diamond and $e^{- L_0 (\tau + L_P)}$ for the larger one.  
Geometrically, at least for geometries conformal to maximally symmetric ones, it would seem that the time evolution operator for the Planck scale time slice between the two diamonds, in DU coordinates, should be $e^{-i L_0 (\tau + L_P)}$.  By Planck scale time slice, we mean one Planck time along the geodesic. Since this operator is a unitary in the larger Hilbert space it will, in general, entangle the smaller diamond's fermions with those of the larger diamond.  

On the other hand, if we think about time evolution between time $\tau - L_P$ and $\tau$ and time $\tau$ , then the evolution operator in that time slice is $e^{iL_0(\tau)}$, but the extra degrees of freedom that are added between $\tau$ and $\tau + L_P$ must remain decoupled during evolution from $\tau - L_P$ to $\tau $ .  Combined with conformal invariance this motivates us strongly to add the new fermions as free fermions.   That is, between time $\tau$ and $\tau + L_P$ the full evolution operator on the entire Hilbert space is 

\begin{equation} e^{-iL_0 (\tau + L_P)} \otimes e^{- iL_{0}^{out} }.\end{equation}
  $L_{0}^{out}$ acts on the tensor complement of the diamond Hilbert space of the interval $[-T,\tau + L_P]$ in the Hilbert space of $[-T,T]$.  The tensor complement can be generated by fermions that are added in each of the intervals $\tau + n L_p$ to $\tau + (n + 1) L_P$. 
Choosing their Hamiltonian to be that of free fermions in $L_{0}^{out}$ guarantees that the time evolution preserves causality and also "prepares" them for the interaction with the expanding nest of prior diamonds.

The interaction, to leading order in $L_P/L$, is largely determined by two constraints: $1+1$ dimensional conformal invariance and {\it fast scrambling}\cite{lshpss}.  The evidence for fast scrambling comes predominantly from the rapid homogenization of perturbations on the horizon of a causal diamond, as viewed from an accelerated trajectory that avoids penetrating the horizon. For non-negative c.c. or for boundary anchored RT diamonds in AdS space, perturbations become homogeneous in times of order the horizon curvature scale.  In\cite{tbwf17} we argued that this could be explained if the dynamics was invariant under area preserving diffeomorphisms in $2$ or more dimensions.  This is a global, rather than a gauge symmetry.  It implies that the dynamics has no respect for geodesic distance in the transverse space.  A small spherical cap is equivalent to a many fingered amoeba of the same area, which touches points that are arbitrarily far away on the manifold.

\begin{figure}[h]
\begin{center}
\includegraphics[width=01\linewidth]{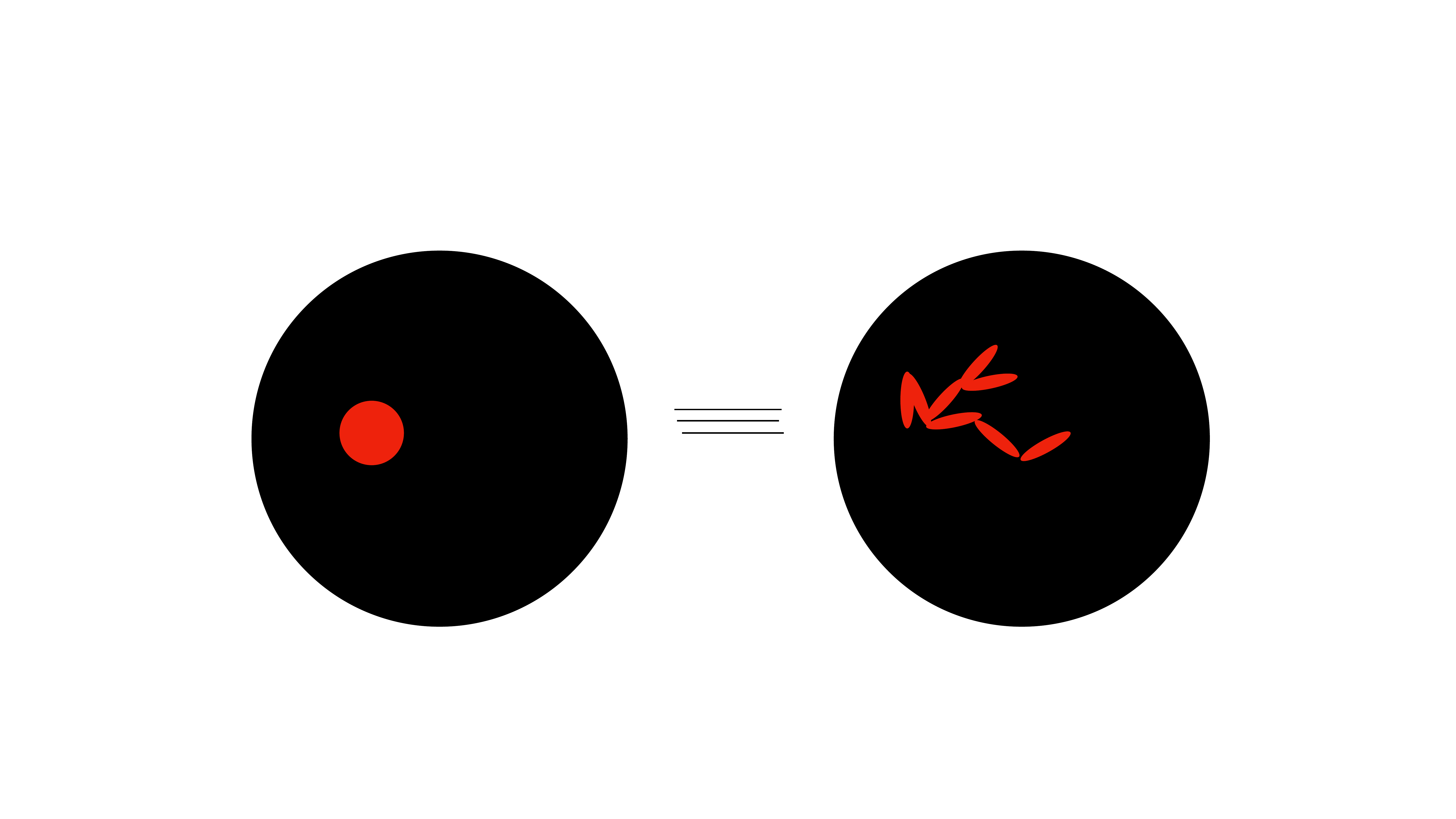}

\caption{Holoscreen dynamics invariant under area preserving diffeomorphisms can propagate quantum information without regard to geometric distance. Consistent with exponentially fast homogenization of charge and mass on horizons.} 
\label{fig:amoeba}
\end{center}
\end{figure}
 Fast scrambling is often defined by the statement that a perturbation of a single q-bit operator at time zero, will fail to commute with every other q-bit in a time that scales like the number of q-bits.  It is characteristic of models that involve "all to all" couplings of a set of q-bits.  
The homogenization of mass and charge in logarithmic time certainly does not require non-locality which is quite so drastic.  A paradox pointed out by Hayden and Preskill\cite{hp} is the strongest argument for the logarithmic time scale.  It's not clear, at least to this author, whether it implies the kind of "all to all" coupling that is often assumed.

Since our fundamental variables are spinors, we can easily construct differential forms
\begin{equation} J_{(m(1)\ldots m(k))}^a = \bar{\psi} (\sigma , x) \Gamma_{(A_{1}\ldots A_{k})}\gamma^a e_{m(1)}^{A_1} \ldots e_{m(k)}^{A_k} \psi (\sigma , x) , \end{equation} 

where $e$ is the veilbein on the transverse manifold, the $\Gamma$ matrices are $d-2$ dimensional Euclidean signature and the $\gamma$ matrices $2$ dimensional Lorentzian signature.  Recalling that our manifold has fixed volume, we can make area preserving diffeomorphism invariant Hamiltonians by integrating products of differential forms of total rank $d-2$.  If the fermions at each $x$ were independent canonical variables then we could construct independent $U(1)$ currents for differential forms of each rank $p$
\begin{equation} J_a^p (x,\sigma) = \bar{\psi} (x,\sigma) \gamma_a \Gamma_{(A(1) \ldots A(p))} e_{m(1)}^{A(1)} (x) \ldots e_{m(p)}^{A(p)} (x) \psi (x, \sigma) . \end{equation}
If $d$ is even the currents with rank $p$ and $d - 2 - p$ commute with each other and we can construct a fixed line of conformally invariant Thirring interactions from products of these currents.  For $p$ odd, we would have to use the reducible representation of the $d - 2$ dimensional Clifford algebra and insert the extra anti-commuting element $\Gamma$ into either the $p$ or $ d - p - 2$ rank current to construct a similar interaction.  These interactions are formally invariant under volume preserving diffeomorphisms.   However, if the $\psi (x,\sigma) $ for different $x$ were truly independent two dimensional fermion fields, the interactions would be ultra-local in $x$ and could not propagate information along the horizon at all.  
It's the fixed cutoff on transverse Dirac eigenvalues that leads to fast scrambling.

This same cutoff will violate $1 + 1$ dimension conformal invariance, but recall that the arguments of Carlip and Solodukhin are only valid to leading order in an expansion in $L_P/L$.  The finiteness of the entropy of the causal diamond already told us that the CFT had to be cut off, violating conformal invariance.   A central conjecture of the current paper is that for large $L/L_P$ the system is fast scrambling, but also close enough to a CFT that the Cardy formula for the spectral density is valid in the vicinity of the value of $L_0$ picked out by the classical argument of Carlip and Solodukhin.

Our proposal now is that we choose the CFT cutoff once and for all by the criterion that for each causal diamond along a nested sequence of proper time intervals the modular Hamiltonian is the $L_0$ generator of the cutoff CFT, where the cutoff is imposed by choosing an $L_0$ value closest to the classical "Liouville" value computed by Carlip and Solodukhin's methods, and one chooses an interval of discrete eigenvalues around that value such that 
\begin{equation} \langle L_0 \rangle = \langle (L_0 - \langle L_0 \rangle)^2 \rangle = \frac{A}{4G_N} . \end{equation}
This condition is enforced for some minimal area diamond, the smallest area for which we believe the CS arguments are valid.  It determines the cutoff on the width of the band of $L_0$ eigenvalues once and for all.
Only the number of fermion fields is allowed to vary as we change the proper time interval along the geodesic, and that changes with the area of the transverse manifold in order to continue obeying the above equation.  The change is implemented by changing the eigenvalue cutoff on the transverse Dirac operator {\it for the appropriate transverse manifold at each time slice.} Time evolution in the nested sequence of diamonds 
is given by $e^{iL_0 (\tau + L_P)}$ in the wedge between the diamond whose future tip is at $\tau$ and the next one in the sequence.  $L_0 (\tau + L_P)$ is the modular Hamiltonian of the empty diamond state of that large diamond.  

This procedure produces a finite quantum model for a given choice of a solution of Einstein's equations with a stress tensor satisfying the null energy condition, at least if the geometry is conformal to a maximally symmetric space-time.  In higher dimension there seems to be a bit of ambiguity in the choice of Thirring couplings for the various allowed values of $p$.  These models are certainly well defined and have the hydrodynamic behavior that Jacobson's argument tells us to expect from Einstein's equations.  By construction they obey a primitive notion of causality:  the (even functions of the) degrees of freedom in a given causal diamond form a factor in the algebra of all operators in the Hilbert space of a large finite proper time interval\footnote{As always, we inserted a proper time cutoff in order to deal only with finite dimensional algebras.  We know how to take the limit for AdS/CFT, and it involves a drastic change in the formalism in which we introduce a tensor network of the systems discussed here in order to obtain a QFT limit.  This is discussed in the appendix. Below we'll discuss the subtleties of the infinite entropy limit in asymptotically flat space.}, and time evolution preserves that factor throughout the proper time in the diamond.  To generalize these models to arbitrary geometries, one would have to find the analog of DU coordinates for each space-time.  That is, the geometric transformation that acts as $L_0 = \partial_t$ on the holographic screen of each diamond, should be the boundary value of a vector field whose flow lines inside the diamond define a set of time-like trajectories (including the geodesic) which start and end at the tips of the diamond.  

The formalism so far is missing at least three different key elements:   it is tied to a particular geodesic and does not obey a "principle of relativity".   The correspondence with local quantum field theory is obscure. The connection to the AdS/CFT correspondence, our best understood model of quantum gravity, is also absent.  We will deal with the last of these issues in an appendix.   The other two will be discussed in the following sections.

\section{The Quantum Principle of Relativity}

Given two proper time intervals along geodesics $G_{1,2}$, the corresponding causal diamonds have an intersection , which might be empty.  The intersection is not itself a diamond, but contains a maximal area diamond, $D_{12}$, unique up to symmetries.  The Quantum Principle of Relativity (QPR) is the statement that $D_{12}$ corresponds to an isomorphism between a tensor factor in the Hilbert space of $D_1$ and that of $D_2$.  Given  initial states on the past boundaries of $D_{1,2}$, there are density matrices $\rho_{1,2}$ on this tensor factor.  The QPR further states that these density matrices have the same entanglement spectra.  The QPR is the dynamical principle that makes the Hilbert bundle over the space of geodesics into a quantum version of Einstein's equations. It has, unfortunately, proven extremely difficult to find efficient machinery for implementing this principle.  What has been done is, I believe, non-trivial, but is mostly at the level of pictures and stories.  

The simplest complete HST model is one in which space-time is a flat Friedmann-Robertson-Walker (FRW) universe with scale factor \begin{equation} a(t) = \sinh^{1/3} (3t/R) . \end{equation} This is a universe that begins its life with equation of state 
\begin{equation} p=\rho , \end{equation} and asymptotes to dS space with Hubble radius $R$.  The quantum model consists of fermions fields $\psi (\Omega,\sigma)$ on an interval $\times S^2$ with an angular momentum cutoff that increases linearly with time.  The fermion Hamiltonian is the sum of a free $1 + 1$ dimensional Dirac Hamiltonian for every transverse angular momentum mode and a term
\begin{equation} g \int d\sigma \ J_a^0 (\sigma, x) J^{a 2} (\sigma, x) ,\end{equation} built from the product of the 0-form and 2-form currents.  The integral includes an integral of the two form over the transverse two sphere. When the expectation value of $L_0 (t_{max}) $ reaches the dS entropy $\pi R^2$, we stop increasing the number of fermion fields and continue the time evolution of the system with $L_0 (t_{max})$ as the Hamiltonian.  

Now consider another geodesic in the same space-time and the causal diamond corresponding to the interval $[0,\tau]$ in proper time along that diamond.  The intersection with the diamond $[0,\tau]$ along the original geodesic contains a maximal diamond of area $A(\tau, D)$ where $D$ is the space-like distance between the two geodesics on some fiducial surface of fixed proper time.  As long as the rule for the dynamics in any diamond is that for the empty diamond initial state, the modular Hamiltonian in any subdiamond is unitarily equivalent to $L_0 (A)$ for the same CFT described above, with the appropriate value of the central charge. Then the QPR will be satisfied, if the initial state is chosen to be such that, for each choice of geodesic, the initial state is a tensor product of $e^{-L_0^{min}}$, where $L_0^{min}$ is the Virasoro generator of the CFT corresponding to the minimal diamond for which the Carlip-Solodukhin analysis is valid.  This is close\footnote{The CS analysis provides a finite quantum model of the space-time down to diamond areas of Planck scale.  The classical equations clearly break down in the vicinity of the Planck scale, while the CS {\it model} does not.  The question is whether there is any reason to believe this particular finite quantum mechanical model.  A physicist's answer should be that only experiment could decide.} to the minimal diamond for which a description of the space-time as a flat FRW model with equation of state $p=\rho$ makes sense.  Once the total entropy gets small enough, the description of the density of states by Cardy's formula will become less accurate.  In our model of the system as fermion fields with Thirring interactions, the detailed values of the couplings and the precise nature of the cutoff will become more important.  Although the quantum model that Carlip and Solodukhin guessed from hydrodynamics is well defined, we have no particular reason to consider it to be the only correct answer when the diamond becomes small.  If one were really probing Planck scale distances in the laboratory, only experiment would be the arbiter of what the right microscopic theory was.  

From the point of view of a space-time picture, we can think of this as a restriction on how useful it is to think of two different FRW geodesics as being spatially close to each other on a fixed time slice.  The initial state above, with a given choice of the minimal entropy, prescribes a minimal initial distance between FRW geodesics.   If evolution along each geodesic is chosen to be identical (this is the definition of our Hilbert bundle model, not a fine tuning of the initial state), then the QPR will be satisfied.   We have described elsewhere\cite{holoinflation} how to construct more interesting, inflationary, cosmologies based on these ideas.  In those models the QPR implies that inflationary horizon volumes along distant geodesics appear in the DU coordinates of a given geodesic as a dilute gas of black holes.  Some mild assumptions about the properties of those black holes produce a very economical theory of early universe cosmology, which accounts for CMB fluctuations\footnote{and makes predictions that are in principle distinguishable from those of QFT models of inflation.}, the Hot Big Bang, baryogenesis, and dark matter.   The model has no Transplanckian problem, and is completely consistent with quantum mechanics, causality and unitarity.

The QPR is also essential for establishing that our Hilbert bundle formulation of quantum gravity actually corresponds to unitary evolution on a Hilbert space.  The prescription that the DU evolution operator on the time interval $[\tau, \tau+L_P]$ is $e^{-i L_0 (\tau + L_P)} $ defines time evolution in terms of embedding maps of the Hilbert space of a smaller diamond into that of a larger one. When we compare two diamonds along the same geodesic, whose future tips differ by one Planck unit, we add new degrees of freedom only along the new section of null boundary.  These degrees of freedom are however already included in the diamonds of smaller proper time, along other geodesics.  So the QPR puts constraints on the evolution of these "out of diamond" degrees of freedom.  Applying the QPR along every pair of geodesics we conclude that on the full Hilbert space along each geodesic there is a unitary operator $U(t) = U_{in} (t) \otimes U_{out} (t)$, where $U_{in} (t)$ operators only on the Hilbert space of the diamond at time $t$, and $U_{out}$ only on its tensor complement.  Strictly speaking, this argument only applies to the dS case, where the Hilbert space of a complete geodesic is finite dimensional.  For non-positive c.c. we must be more careful about infinite dimensional limits.  We'll discuss some aspects of this in the next section and in the appendix.  Above we have proposed that $U_{out}$ be defined in terms of the $L_0$ of a collection of free fermion fields.   An important question that we have not yet studied is whether this proposal is consistent with the QPR.  The real force of the QPR comes into play for states describing localized excitations in a diamond, to which we  now turn.

\subsection{Space-time Locality and the Origin of Quantum Field Theory}

The formalism described so far allows us to localize quantum information on the boundaries of causal diamonds in space-time but has not given us any clue about the description of the quintessential object of experimental physics: a localized excitation traveling on a time-like or null trajectory through space-time, particularly a timelike geodesic.  The essential hint about the description of localized objects came from the study of black holes in de Sitter (dS) space.  Before explaining that, we should comment on the unfortunate fact that localization in AdS space is quite a different story, which we will mostly relegate to an appendix.  In brief, AdS space is best described by the Error Correcting Code/Tensor Network formalism, in which a cutoff $1 + d - 2$ dimensional conformal field theory is described as a sequence of lattice field theories arranged as shells in $d - 1$ dimensional space, with embedding maps (called a Tensor Network Renormalization Group) connecting consecutive shells in the sequence\cite{swingleetal}.  This is a manifestly local presentation of an approximation to $AdS_d$ geometry, which also manifests Maldacena's scale radius duality.  
When the radius of AdS space is much larger than all microscopic scales, the number of degrees of freedom inside each node of the tensor network is very large and space-time locality has to do with what is going on inside the node, rather than with communication between the nodes.  The ECC/TN picture becomes irrelevant.  Thus, our claim is that locality on scales smaller than the AdS radius is a property of the cutoff lattice models of the tensor network, and cannot be addressed easily by computations in the continuum boundary field theory.

In the year 2000, Fischler and the present author independently conjectured\cite{tbwfds} that the finite Gibbons-Hawking entropy of dS space implied that the Hilbert space of a quantum model of dS space was finite dimensional.  An essential part of that conjecture was the fact that there is a maximal mass black hole in dS space, so that it is extremely plausible that a quantum model of dS space has a UV cutoff.  In discussing this at the Festschrift for L. Susskind at Stanford, Fischler and I were struck by the fact that the Bekenstein-Hawking area law for black holes much smaller than the maximal size in dS space showed that the total entropy of the space-time with a black hole in it was {\it less} than the entropy of empty dS space!  Moreover, this gave an almost classical derivation of the Gibbons-Hawking temperature of dS space, which Gibbons and Hawking had derived from quantum field theory in a de Sitter background\cite{GH}.  We conjectured that the density matrix of empty dS space was maximally uncertain (and that the same was true for any empty causal diamond of finite area\footnote{A similar claim was made by R. Bousso\cite{bousso} at about the same time.}) and that energy in the static patch was {\it defined} by saying the number of "constraints" on the states was of order $ER$, where $R$ was the dS radius.   We did not have a very clear model of this in mind but in the summer of 2006 Tomeu Fiol walked into my office in Santa Cruz and presented me with such a model.

Suppose one has a matrix, with matrix elements that are fermionic oscillators
\begin{equation} [\psi_{i}^j, \psi^{\dagger\ k}_l]_+ = \delta_i^k \delta^j_l . \end{equation}   Then if we insist on being in a subspace of the Hilbert space of fermions where we set the off diagonal blocks between a block of size $n$ and one of size $N$ to zero, the statistical probability of being in a quantum state that has order one probability of being in that subspace is
$P \sim e^{- n N}, $ which looks like a Boltzmann factor, with $N^{-1}$ proportional to the temperature and $n$ proportional to the energy.  Something we did not realize at the time, was that if we only require this when the probability is small, then it remains true for a wide variety of density matrices which are not maximally uncertain\cite{tbpd22}.  In particular, even if dS space follows the Carlip-Solodukhin law,  the explanation of dS temperature can be valid for the states in dS space that, according to CKN, are well described by QFT. 

\begin{figure}[h]
\begin{center}
\includegraphics[width=.95\linewidth]{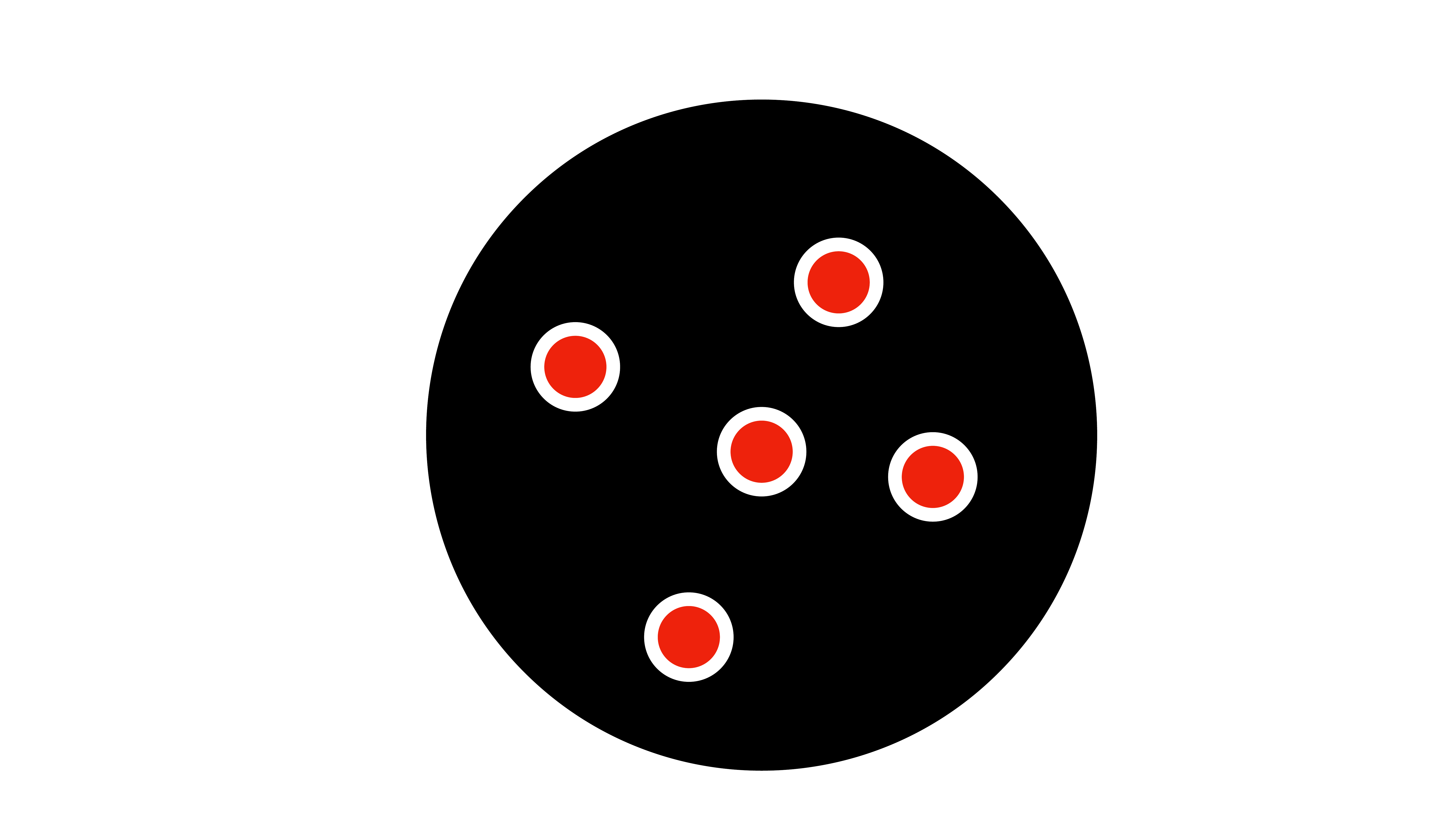}

\caption{The holoscreen of a diamond with localized excitations penetrating it. Red are regions where jet momentum flows into or out of the diamond. White regions have strictly zero momentum, while black regions have momentum flows of magnitude less than the inverse diamond radius.} 
\label{fig:locstat}
\end{center}
\end{figure}

Fiol's matrix model idea needs a little tweaking to make it compatible with the ideas presented in the present paper. Some of that tweaking was already done in 2006\cite{bfm}.  First of all, the fermions should be spinors.  In four dimensions, we can think of the cut-off spinor bundle as $N \times N + 1$ matrices, since these can be thought of as transforming in all half integer spin representations of $SU(2)$ up to $J = N - \frac{1}{2}$. We can then construct square $N \times N$ matrices $M_{i}^j = \psi_i^A \psi_A^{\dagger\ j}$ and impose the constraints on states in terms  of these.  These bilinears in spinors can be thought of as "fuzzy differential forms on the two sphere".  The trace of products of these matrices is the fuzzy analog of the integral of two forms over the sphere.  When a matrix is block diagonal its trace is the sum of the traces of the blocks.   We can think of this as writing the sphere as a line bundle over an interval and doing a one dimensional sum over slices of two volume which are then discretized into matrices with matrix elements that are bounded operators in a fermionic Hilbert space.  Block diagonal matrices correspond to bands of vanishing volume form.
This does not quite capture the full structure of a trace however, since we can do unitary transformations, which change the order of the blocks.  A better picture then is regions of finite area surrounded by bands where the variables vanish.  \ref{fig:locstat} is a cartoon of what we mean.  

 Readers will note the large black region on the figure, the white bands of vanishing variables, and the smaller red regions.   The latter represent small blocks of the matrix of size $n_i$ while the black region has size $N - \sum n_i \gg 1$.  Constraints setting the off diagonal blocks between different small blocks to zero can easily be removed by finite time dynamics, while it's plausible that with an appropriate scaling law for the Hamiltonian, the $o(N)$ constraints between the large block and each of the small ones will never be removed.  In the strict $N\rightarrow\infty$ limit, $\sum n_i$ would become an asymptotic conservation law if the proper time was also scaling like $N$.   

On the other hand, in our model for dS space, $N$ is kept finite while the proper time goes to infinity, so the quantity $\sum n_i$ is not conserved, but controls the Boltzmann factor of low probability states in the density matrix.   $\sum n_i$ is thus proportional to the static energy, which will become the conserved Minkowski energy if we let $N$ go to infinity with proper time.  The measure theoretic cartoon of \ref{fig:locstat} tells us how to generalize these ideas to higher dimensions, for large $N$.  Isolated localized excitations entering a causal diamond with area large in microscopic units, but small compared to the scale defined by the c.c,, are defined by red regions of finite area $A_i$ on the holographic screen, whose area scales like $N^{d-2}$.  The energy carried by those excitations is proportional to $A_i^{\frac{d-3}{d-2}}$, {\it i.e.} the volume of its bounding surface.  The empty region surrounding 
the red region has an area $\sim N A_i^{\frac{d-3}{d-2}}$.  To implement this picture in a matrix model, we need to write the matrices as 
\begin{equation} M_i^j = \psi_i^A \psi^{\dagger\ j}_A , \end{equation} where the index $A$ runs over of order $N^{d-3}$ values.  Fischler and the author\cite{tbwfnewton}, using the fact\cite{tbjk} that the number of solutions of the Dirac equation with angular momentum cutoff $N$ is equal to the number of components of a $d-2$ form in $N$ dimensions, tried to write $\psi_i^A$ as a $d-2$ form.  However, one would also like to be able to impose constraints on of order $n_i^{d-3} N$ fermion number operators in order to make all the off diagonal matrix elements between $n_i$ and $N - n_i$ indices vanish.   It is not yet clear how to do this.  Alternatively, we could try to implement the constraints directly from the figure in terms of the "local" spinor field expanded in a basis of solutions of the Dirac equation with an eigenvalue cutoff.  Fuzzily localized fields cannot reproduce exact characteristic functions of regions but for large values of the cutoff the approximation should be good enough for statements about non-interaction of the localized objects with the horizon to be valid.  Indeed, a Hamiltonian written in terms of integrals of products of differential forms constructed from truly local fermions would be ultra-local and could not propagate information around the horizon.  Fuzzy localization is essential if we are to reproduce the fast scrambling properties of horizons.   In four dimensions there is a well known connection between the matrix model formulation and area preserving diffeomorphisms, with unitary conjugations of the matrices playing the role of a regularization of the infinite dimensional group of area preserving diffeomorphisms.  We need a d dimensional generalization of that formalism.  

It is unlikely that there is an interesting quantum theory of dS space in dimension greater than $4$.  Since 2001\cite{tbwf99}, Fischler and the present author have argued that dS space should be thought of as a regulated version of a scattering theory of "particles" in Minkowski space\footnote{The reason for the scare quotes will become clear in the next section.}.  There is no S-matrix in dS space:  every state returns to a maximal entropy equilibrium with no local excitations in a time in any static coordinate system which is not much longer than the dS radius\footnote{This statement is irrelevant for sufficiently complex objects like a galaxy.  The constituents of the complex object decohere its center of mass trajectory and define a classical reference frame independent of any mathematically defined coordinate system.} .  However, if the $R_{dS}/L_P \rightarrow\infty$ limit of finite time transition operators in dS space defines a Minkowski scattering operator, then we can view this as a mathematical definition of dS space that is independent of particular complex localized objects (see previous footnote) .  
However, we have an enormous amount of evidence from string theory that Poincare invariance implies exact super Poincare invariance in models of quantum gravity.   Rigorous theorems in classical SUGRA show us that there are no dS solutions of SUGRA for $d \geq 5$.  

\begin{figure}[h]
\begin{center}
\includegraphics[width=01\linewidth]{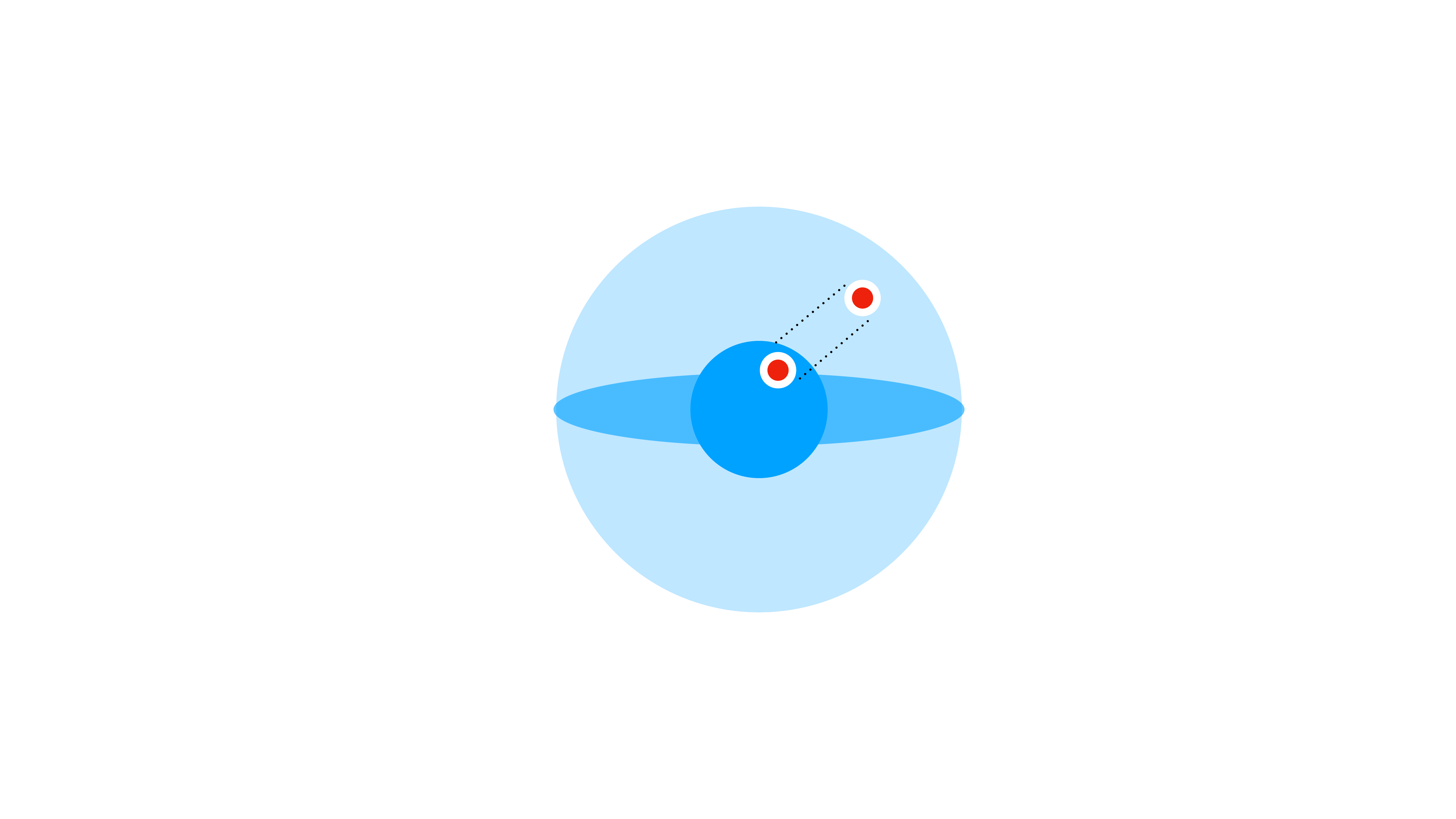}

\caption{The holoscreens of two consecutive nested causal diamonds, showing how following the constraints leads to a picture of the trajectory of a jet of particles through space-time. } 
\label{figure:layers}
\end{center}
\end{figure}

The real reason to be interested in this description of localized excitations for $d > 4$ is that it applies equally well to the Minkowski limit.  Indeed it provides the crucial link between the formalism we have been discussing and QFT.  Let us begin by noting the analogy between \ref{fig:locstat} and a layer in a modern particle detector.  We can think of the red regions as the pixels in the detector that have been "lit up" by a particular scattering event and the white regions as "jet isolation zones" that are used to define particle jets.  In a real detector these are defined by including all soft particles that we think should be included in the jet, leaving an empty region around it.  The dark regions are pixels that don't light up, which means that anything that flowed through them was below the detector threshold.   Here we are insisting that not even "asymptotically zero energy" flows through the annular regions surrounding the jets.  The dark region has energy of order $1/N$ as measured along the geodesic in a finite area diamond of area $N^{d-2}$ in Planck units, Now imagine a time symmetric nesting of diamonds as a very very finely layered particle detector.   Then following the red regions from layer to layer, we get a picture of the trajectory of the jet through the bulk of space-time\ref{figure:layers}.
  In the description of this trajectory from the point of view of the Hilbert fiber over a given geodesic, all we learn about this trajectory is how much area it takes up on each holoscreen in the nest of diamonds, but the QPR relates this to the Hilbert fibers over all other trajectories and consistency between them determines the geometric shape of the jet trajectory through space-time.  

\begin{figure}[h]
\begin{center}
\includegraphics[width=01\linewidth]{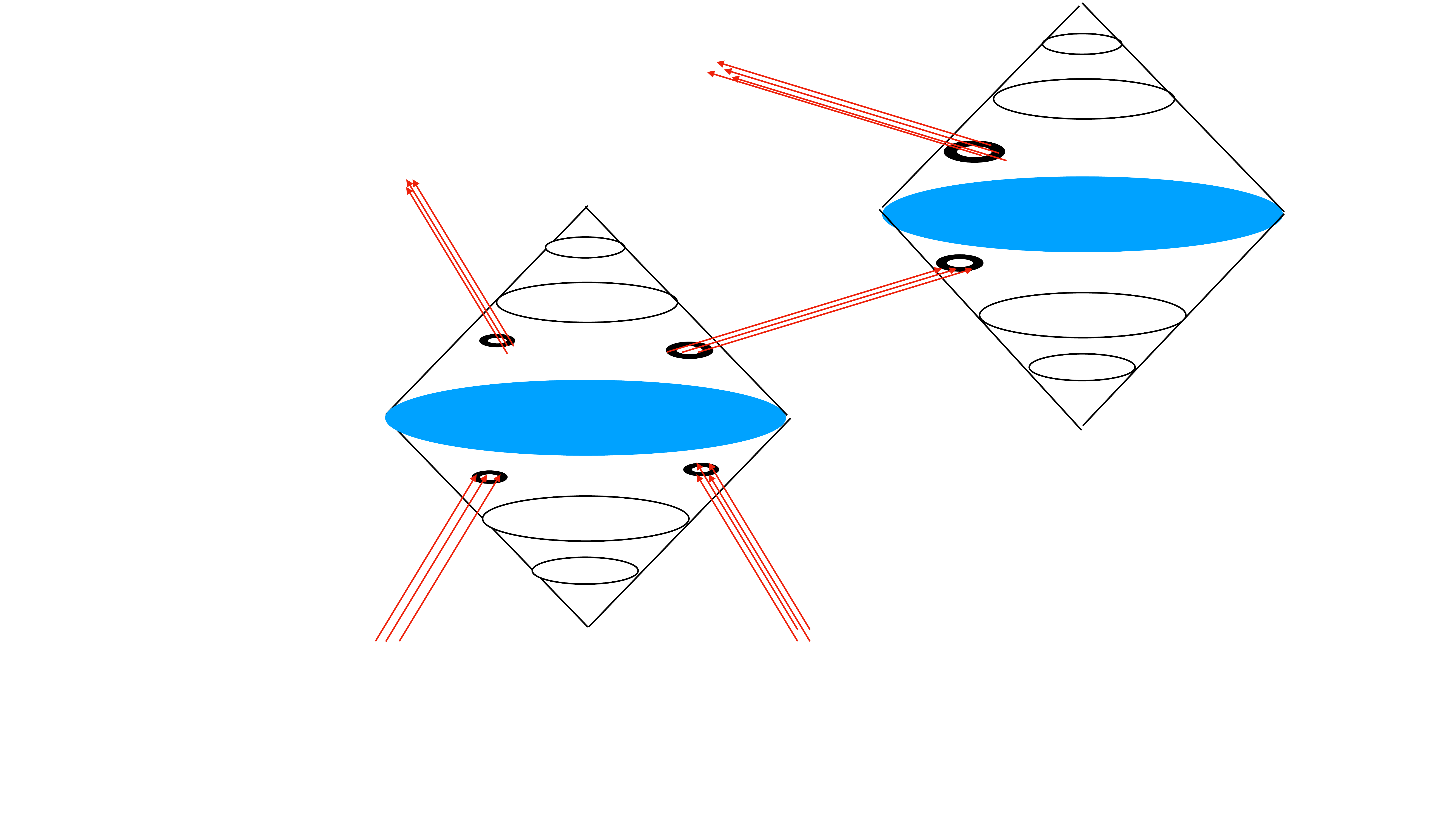}

\caption{Decomposition of amplitudes in HST models into time ordered Feynman-like diagrams, describing jets of particles propagating between causal diamonds in the background space-time. } 
\label{fig:feynman}
\end{center}
\end{figure}

Thus, given an initial state on a very large diamond, containing some number of well separated jets on its past boundary, we can compute the amplitudes for Feynman like diagrams describing their evolution into a different collection of well separated jets on the future boundary (\ref{fig:feynman}).  

These are not exactly Feynman diagrams because each picture in which energy $E$ propagates between two points contains amplitudes for this energy to split up into distinct pieces carrying fractions of the total energy.  The reason for this is that each jet contains all fermion operators contributing to the matrix $M_{i}^j$ with indices running from $1$ to $n$ with $n^{d-3}\propto E$.  

It is also very easy to see where the particle/jet picture characteristic of QFT breaks down in these models.  Suppose that for some macroscopically large $N$ we find that $\sum n_i^{d-3} \sim N^{d-3} $ .   Then clearly the approximation that we have a small deviation from the equilibrium state has broken down, and we can expect a rapid evolution back to equilibrium.  This is precisely the criterion that the Schwarzschild radius is of order the size of the diamond, here derived from a quantum model, rather than general relativity.  

There are other ways in which the diagrams derived from the Hilbert bundle formalism differ from Feynman diagrams.  They are definitely causally ordered.  The meaning of a diagram like that of figure \ref{fig:feynman} is that a jet that exits the future boundary of the diamond $1$ enters the past boundary  of diamond $2$ and interacts with other jets entering that diamond, producing some final state, for which the diagram computes the amplitude\footnote{Of course another big difference is that we do not have a nice closed formula for the amplitude, just a sort of Trotter product formula in terms of amplitudes in $1 + 1$ dimensional CFTs with time varying central charge. }.   The resemblance to time ordered Feynman diagrams then is that these pictures give a bulk space-time decomposition of an amplitude into processes that occur in different causal diamonds.  Unlike Feynman diagrams they include both small diamonds, which resemble localized vertices in which a number of particles come into a diamond and a generally different number emerge and "black hole formation processes" in which particles enter a diamond, which then retains its identity as a meta-stable object of fixed energy, emitting jets of energy in a random thermal manner for an extended period.  As far as scaling laws are concerned, these objects behave like the black holes of general relativity, but are excitations of a unitary quantum mechanical system.    In the next section, we will indicate how these models can lead to a Lorentz invariant scattering theory in Minkowski space.

\section{Representation of Non-compact Isometry Groups}

All of the maximally symmetric space-times have non-compact groups of isometries, which we will denote collectively by $G$.   Finite causal diamonds in these space-times are invariant only under an $SO(d-1)$ subgroup and geodesics only under $SO(d-1) \otimes R$.  Transformations in $G/(SO(d - 1)\otimes R)$ map one fiber of the Hilbert bundle onto another and so are not in general represented on any fixed Hilbert space.  The static time translation $U_S$ maps one interval along a geodesic to another one.  It is {\it not} the same as the time dependent Hamiltonian that describes propagation in a diamond.  In terms of conventional coordinate systems $U_S$ uses the global time slicing, while the Hamiltonian $H(t) = i U(t + 1) U^{\dagger} (t) - 1$ describes propagation between slices within the diamond, as in Figure 1.  For maximally symmetric space-times these are the {\it diamond universe coordinates} of\cite{CHM}\cite{JV}. In the case of dS space, $H(t)$ approaches the global Hamiltonian $i \partial_{t^{s}}$ as the static time goes to infinity. The asymptotic dimension of the Hilbert space is $e^{A_{d-2} R_{dS}^{d-2} /4G_N}$ up to subleading corrections in the exponential.  $A_{d-2}$ is the area of the unit sphere.  Strictly speaking, the area is just the expectation value of the modular Hamiltonian, but in\cite{tbpd22} we've shown that the conditions $\langle K \rangle \rightarrow {\rm ln\ dim} {\cal H}$ and $(\Delta K)^2 = \langle K \rangle$ are compatible in the limit of a Hilbert space of large dimension.  

For AdS space, as $t$ approaches $\pi R_{AdS}/2$ the Hilbert space dimension goes to infinity and the Hamiltonian approaches the generator of the AdS isometry group that leaves a particular time-like geodesic invariant.  The AdS/CFT correspondence tells us that this is the Hamiltonian of a CFT, and the c.c. is determined by the asymptotic density of states in that CFT.  The transformations that do not leave that particular geodesic invariant act as unitary transformations on the CFT Hilbert space.  So in this case, the full isometry group acts on a single Hilbert space.   We can think of the CFT Hilbert space as being a trivial Hilbert bundle over the space of time-like geodesics in AdS space.  A choice of the Hamiltonian $K_0 + P_0$ within its conjugacy class in the conformal group is equivalent to a choice of timelike geodesic and we can think of the fiber over that geodesic as the Hilbert space expressed in the basis where $K_0 + P_0$ is diagonal.  There is no holonomy when we traverse a closed loop in the conjugacy class, so the bundle is trivial. We'll discuss the case of negative c.c. in more detail in the appendix.  

The case of vanishing c.c. is more subtle and is intertwined with the existence (or not) of the Bondi-Metzner-Sachs algebra on the Hilbert space on which the S-matrix acts.  There is also considerable confusion about whether that Hilbert space is a traditional Fock space.  It is clear that it is not in four dimensions, and the present author has argued for some time\cite{tbir} that Fock space is not the correct arena for asymptotically flat quantum gravity in any dimension.  That conviction grew out of the picture of particle/jets as constrained states of holographic degrees of freedom on the boundaries of a diamond, which we described in the previous section.  In the appendix, we'll see that the same conclusion follows if we follow the approach of Polchinski and Susskind\cite{polchsuss} and derive the Minkowski S matrix as a limit of CFT correlators.

Let us then consider the variables in a finite causal diamond in $d$ dimensional Minkowski space times a compact $11 - d$ manifold ${\cal K}$ of fixed size much larger than the $11$ dimensional Planck scale.  The magic number $11$ appears because we anticipate the result of our investigation, namely that the limiting theory must be super-Poincare invariant.  Decades of investigation into string theory make it fairly clear that every supersymmetric model of quantum gravity can be viewed as a compactification plus duality transform of eleven dimensional SUGRA.  

The variables are fields $\psi_a^{I,M} ( {\bf \Omega}, \sigma)$ .  $a$ is a two dimensional Dirac index and $I$ a $d-2$ dimensional Dirac index, while $M$ labels a finite set of eigenfunctions of the Dirac operator on ${\cal K}$.  We will omit the indices $I,M$ in future equations.
  ${\bf \Omega}$ is a coordinate on the $d - 2$ sphere.  The spherical harmonic expansion on this sphere is cut off, but we're considering the limit of an infinite diamond where the cutoff goes to infinity.  Recall that the volume of the sphere scales like $N^{d-2}$, where $\pm N $ is the scaling of the cutoff on the Dirac eigenvalue.  

Now we want to define in(out) scattering states.  We consider a sequence of diamonds with $N_i$ going to infinity, in a future(past) oriented nested covering of the Penrose diagram of Minkowski space (Figure 1).  Pick a finite set of points ${\bf \Omega}_k$ on the sphere.  Around each point we impose a constraint that the variables $\psi_a ({\bf \Omega}, \sigma) $ "vanish" in a shell of volume $N_i A({\bf \Omega}_k)^{\frac{d-3}{d-2}} \equiv N_i E_k $ surrounding an area $ A({\bf \Omega}_k)$, which includes that point.  By independent area preserving maps on each diamond boundary, we can turn each of these regions into a spherical cap surrounded by an annulus of "vanishing" variables.
Of course, with only a finite number of spinor spherical harmonics, we cannot make a function that vanishes exactly in an annulus.  We mean the closest approximation to such a half-measure with the available basis set.

For finite $N_i$ the "areas" are discrete numbers, but in the limit they become continuous and the $E_k$ are continuous variables, which we associate with the magnitudes of null momenta in Planck units\footnote{Stable massive particles are always associated with BPS states or bound states of them and require a generalization of the SUSY algebras we will write below.}.  In other words, as we let $N_i \rightarrow \infty$ we are free to rescale the $E_k$ at will since we can absorb the rescaling into $N$.
As usual, in order to obtain finite results in the limit, one must impose some kind of scaling symmetry, and the obvious one is the conformal group of the $d-2$ sphere, which is the Lorentz group of the null cone $P^2 = 0$ in $d$ dimensions.   Our spinor variables thus become spinors $Q_{\alpha} (P)$ on the null cone and they satisfy the Lorentz covariant condition
\begin{equation} P_m (\Gamma^m)_{\alpha\beta} Q_{\beta} (P) = 0 . \end{equation} which reduces the number of components to the transverse ones that existed before we took the limit.  
In the limit, the $Q_{\alpha} (P)$, are generalized functions on the null cone.  For non-zero values of $P$, these generalized functions only exist in a finite number of spherical caps, surrounded by annuli in which $Q_{\alpha} (0) = 0$.

 Note that we have dropped the two dimensional structure that was a crucial part of the description of the boundary of finite causal diamonds, which we've emphasized by also changing the names of the operators.  That structure was used by Carlip and Solodukhin to understand the finite boundary entropy and by\cite{bz} to understand the finite fluctuations of that entropy.  In the limit, both of these quantities become infinite and must be "renormalized away" if there is to be a finite limiting scattering operator.  We have interpreted the fluctuations implied by this two dimensional structure as fluctuations in the transverse geometry, so its disappearance in the asymptotic limit is the statement that the geometry at null infinity is frozen.
 This implies an asymptotic decoupling of the dynamics of the emergent zero energy "topological" sector of the theory in the infinite diamond limit.  The various "soft theorems"\cite{soft} in the extensive literature on BMS symmetries prove this kind of decoupling, but there is as yet no clear picture of what the Hilbert space of asymptotically flat quantum gravity looks like.  Above four dimensions, most authors would just claim that it is Fock space, but this is an incredibly ambitious claim.  Initial states consisting of a few particles with large sub energies, can create large black holes which orbit around each other for extended periods of time, merge, emit Hawking radiation and so on.   To claim that the final state of any such process is a normalizable state in Fock space, even in 11 dimensional SUGRA where there are no apparent infrared problems in perturbation theory, is the height of hubris.   In the BFSS matrix model\cite{BFSS}, which is a manifestly unitary scattering theory with the particle content of $11$ D SUGRA (in the large N limit), the claim is equivalent to having control over processes in which large block diagonal matrices come in from remote regions of the moduli space and go out into regions in which other large blocks plus an arbitrary number of small blocks with transverse momenta of order $N^{-1/2}$ are emitted.  In four dimensions we {\it know} that Fock space is not the right answer and there is recent evidence\cite{Waldetal} that the Fadeev-Kulish proposal does not work either.  

I have proposed\cite{tbir} that the proper framework for scattering theory in quantum gravity is a map between representations of the algebras of the $Q_{\alpha}^M (P)$ operators on the positive and negative energy parts of the null cone.   These algebras are
\begin{equation} [Q_{\alpha}^{M\ \pm} (P), Q_{\beta}^{N\ \pm} (P^{\prime}]_+ = \pm \delta (P\cdot P^{\prime}) \delta^{MN} P_m (\Gamma^m)_{\alpha\beta} .\end{equation}
The angular delta function follows from the fact that the fermion fields corresponding to individual angular momenta anti-commute. The delta function in the "internal" indices is achieved by taking appropriate linear combinations.  The rest follows from Lorentz invariance and the fact that the objects we are describing are localized in angle.  They cannot carry tensor charges at infinity.  

The delicate question of what the Hilbert space of asymptotically flat quantum gravity is, has no experimental relevance.  For practical purposes, the inclusive cross sections defined by Weinberg are all we could ever hope to measure if we lived in asymptotically flat space, even in four dimensions, where IR effects are most important. In eleven dimensions these IR questions could effect scattering amplitudes only via terms exponential in ratios of Mandelstam invariants to the Planck mass.  In ten dimensional superstring theory they might be of order $e^{- g_S^{-2}}$, exponentially subdominant compared to the leading non-perturbative $e^{- 1/g_S}$ contributions.

We should note that while Lorentz invariance seems necessary in order to get any kind of sensible infinite diamond limit, it also follows from the QPR.  Our formalism has a built in requirement that everything must be invariant under passage to the fiber over any geodesic of the background space-time.  We've described the infinite diamond limit along a particular geodesic, and discovered an apparent need for Lorentz invariance in the infinite diamond limit, which is consistent with the requirement that the scattering operator should be identical to that obtained by working over a different geodesic.

Something similar happens for space and time translation invariance.  Our Hamiltonian description of time evolution in a causal diamond uses a time dependent Hamiltonian because it is evolving in DU coordinates.  However, when we define asymptotic energy in terms of constraints on initial conditions on the past boundary of the diamond we can see that it is asymptotically conserved.  The argument has two parts and its form depends on whether we use the past or future directed nest of diamonds to cover the Penrose diagram.  Let us consider the future directed nesting.  Then, at early times it is unlikely that the constraints are imposed on the degrees of freedom in the small nested diamond.  Those outside the diamond are described as free fermions, so the constraints are preserved by time evolution.  However, even after the constrained DOF are engulfed by the growing diamond, the size of the Hamiltonian is of order $1/R_{diamond}$, the Hamiltonian is $4-local$ in the fermion variables, and the number of constraints is of order the radius of the Penrose sphere, which we are taking to infinity.  The asymptotic energy is proportional to the coefficient of that infinite number of constraints, so it cannot be changed.  
The same conclusion follows from the QPR applied to causal diamonds that are shifted by a rigid time translation along the same geodesic.

The situation for conservation of spatial momentum is a bit different.   Here we use the QPR to impose constraints on the tensor complement of the Hilbert space corresponding to proper time interval $[-T, \tau]$ along a particular geodesic $G_1$ in Minkowski space in terms of the constraints on the Hilbert spaces corresponding to that same time interval along geodesics related to $G_1$ by a rigid space translation.  In words, these relations tell us when and where a certain amount of energy enters into the causal diamond of a detector traveling along $G_1$.  Thus, although the constraints themselves are defined only in terms of areas, the QPR and the background geometry tell us about the trajectories followed by the centers of jets of particles through the background space-time.  One must recall that the jets are not particles, but actually systems that have (asymptotically) infinite dimensional Hilbert spaces\footnote{For finite size causal diamond the jet Hilbert spaces always have dimension exponentially smaller than that of the diamond, because of the bound that says that the jet energies must be much smaller than the mass of a black hole that would fill the diamond.  But as the diamond approaches infinite size, this allows the jet Hilbert spaces to grow without bound.}.  The jet trajectory is a decoherent semi-classical observable of these large quantum systems. 

Using the area preserving diffeomorphism invariance of the large $N$ Hamiltonians, we can always view the jets as living in spherical caps, surrounded by symmetric annuli where the spinor variables vanish.  The map from the past boundary to the future boundary is invariant under relative rotations of the centers of the jets, since this is a residual symmetry of the area preserving group after we have fixed the jet profiles to be symmetric.  Translation invariance then follows from Lorentz invariance and energy conservation.   It can be derived independently by insisting that the scattering operator obtained at one fiber of the Hilbert bundle be independent of the fiber.  As in the case of energy conservation and boost invariance, the first derivation involves dynamics of the time evolution operator for a given geodesic, while the second is a symmetry constraint on the Hilbert bundle as a whole.  

While these arguments are not rigorous, their conclusion is remarkable, and explains an empirical fact.  What we have found is that the zero c.c. limit of our Hilbert bundle theory of dS space, has stable localized excitations.  If we enforce the Poincare symmetry of Minkowski space, the asymptotic algebra automatically has the form of the supersymmetric generalization of the Bondi-Metzner-Sachs algebra (actually its Fourier transform) derived from SUGRA by Awada, Gibbons and Shaw\cite{ags}. Since the invention of string theory in the late 1960s, every attempt to formulate a non-supersymmetric theory of quantum gravity in asymptotically flat space has met with failure.   If we accept the premise of this paper, that quantum gravity is a Hilbert bundle over the space of timelike geodesics on its hydrodynamic space-time manifold, and the the variables of quantum gravity are Connes-Carlip-Solodukhin conformal fields on causal diamond boundaries, then we understand that failure.   

\subsection{Summary}

The Hilbert bundle formulation of HST resolves all problems with representations of non-compact isometry groups on finite dimensional Hilbert spaces.  In AdS space, the proper time $\rightarrow \frac{\pi R}{2}$ limit, leads to a trivial Hilbert bundle and a unitary representation of the isometry group on the Hilbert space of the CFT on $R \times S^{d-2}$.  In asymptotically flat space we have suggested that one obtains a unitary representation of the semi-direct product of the Lorentz group and the AGS superalgebra (or its generalization to include massive BPS states), but the nature of the asymptotic Hilbert space remains unresolved.  In dS space, which probably only makes sense in dimensions below $5$\footnote{See\cite{satbpdwf} for a model of $dS_3$.}, the isometry group acts only on the Hilbert bundle as a whole, not on any single Hilbert space.  

\section{The 't Hooft Commutation Relations}

The basic postulate of our approach to quantum gravity is that the Einstein equations are the hydrodynamic equations of a collection of quantum systems living on the boundaries of causal diamonds in space-time, whose equilibrium density matrix $\rho = e^{-K}$ satisfies $\langle K \rangle = \langle (K - \langle K \rangle)^2 \rangle = \frac{A}{4G_N}$. When we use the Verlinde analysis and examine the near boundary limit of a causal diamond, the EH action splits into two parts.  Our discussion so far has dealt with the large part of the action, when the transverse size of the diamond $L$ satisfies $L \gg L_P$.  We have argued that it is given by a cut off quantum field theory of $1+1$ dimensional fermions, which is a particular kind of abelian Thirring model.  The fermion fields encode fluctuations of the transverse geometry around a classical background metric $h_{mn}$.   

The authors of\cite{V291} showed that the small term in the action was purely topological (once one imposed the classical equations of motion of the large term), and that its entire content was the 't Hooft commutation relations
\begin{equation}   [X^+ (\Omega), X^- (\Omega^{\prime})] = (\frac{L_P}{L})^{d-4} (\triangle_h - R_h)^{-1} (\Omega,\Omega^{\prime})     . \end{equation}
Our purpose in this section is to provide a model for these commutation relations in terms of the $1+1$ CFT that we have constructed to describe the transverse fluctuations\footnote{For another approach to the 't Hooft commutation relations, see\cite{VZ3}. The first of these papers gives a derivation of $(\Delta K)^2 = \langle K \rangle$ directly from the 't Hooft commutation relations.}  The variables $X^{\pm}$ are defined on the future/past boundary of the causal diamond, and the equal time surface on which the commutation relations are evaluated is the bifurcation surface of the diamond.  The reason that there is such a relation at every point along the past/future boundary is because every point is on the bifurcation surface of some diamond in a future or past directed nested cover of the diamond.  

Our model of transverse dynamics has separate CFTs on the past and future boundaries of a diamond, related by the time evolution operator.  Note that in going from past to future boundary, the role of the time and space coordinates of the CFT are switched.  Thus, the space and time components of a $U(1)$ current are also exchanged.  The space and time components of a conserved $U(1)$ current in $1+1$ CFT have a c-number Schwinger term in their commutator\footnote{A previous relation between the 't Hooft commutation relations and Schwinger terms in 1+1 dimensional electrodynamics was pointed out in\cite{peet}.}.  Thus we postulate
\begin{equation} X^+ ({\bf \Omega}) = \int d\sigma f^+ (\sigma) J^0 ({\bf \Omega}, \sigma) . \end{equation} 
\begin{equation} X^- ({\bf \Omega}) = \int d\sigma f^- (\sigma) J^1 ({\bf \Omega}, \sigma) . \end{equation} 
Here
\begin{equation} \int d\sigma f^+ \partial_{\sigma} f^{-} = 1, \end{equation} and the Schwinger term is
\begin{equation} [J^0 ({\bf \Omega}, \sigma), J^1 ({\bf \Omega^{\prime}}, \sigma^{\prime})] = (\triangle_h - R_h)^{-1} ({\bf \Omega},{\bf \Omega^{\prime}}) \partial_{\sigma} \delta (\sigma - \sigma^{\prime}) . \end{equation}  $\triangle_h$ is the scalar Laplacian of the transverse manifold, and $R_h$ its scalar curvature. 
To understand how to obtain the last equation, we consider a set of $1+1$ dimensional fermion fields labelled by the eigenspinors of the transverse Dirac operator $D$ and write
\begin{equation} J^{\mu} = \bar{\psi} \gamma^{\mu} D^{-1} \psi . \end{equation}

There are two ways in which this formula deviates from that of 't Hooft.   First, the currents, and thus the light ray operators $X^{\pm}$ are bilocal rather than local operators on the transverse manifold.  The second is that the transverse integral operator we obtain from this ansatz is
\begin{equation} D^{-2} = (\triangle_h -  \frac{1}{4}R_h)^{-1} . \end{equation}    The discrepancy in the coefficient of $R_h$ can perhaps be fixed by choosing a non-Riemannian connection for the Dirac operator, though one would clearly like to have a deep physical justification for such a replacement.  Another possibility, which should be explored is to replace the Dirac operator by the Rarita-Schwinger operator as the proper representation of transverse geometry.  If we do choose such a connection, or if the Rarita-Schwinger idea works, then the equations of motion\cite{V291} show that the null energy fluxes, which enter into actual scattering amplitudes are still local operators. The very preliminary nature of this analysis should be evident.   

\section{Conclusions, Speculations and (Ugh) Philosophy}

For many researchers in quantum gravity there is much in the present work that will appear negative.  Quite frankly, it is probably because some of the conclusions I've been drawn to were so distasteful that it has taken a long time to come to the present formulation of HST, despite the fact that they were implied by the basic postulates of the formalism.  I've long believed that the deepest insight into the problem of quantum gravity was Jacobson's demonstration that Einstein's equations, with an arbitrary stress tensor satisfying the null energy condition, were the hydrodynamic equations of the Bekenstein-Hawking entropy law, applied to the boundary of an arbitrary diamond in space-time.  If one thinks about the role the Navier-Stokes equations play in the physics of ordinary matter, Jacobson's paper should have made us all extremely suspicious of the idea that quantum gravity was all about quantizing Einstein's equations.   Extremely suspicious as well of the idea that there was some kind of "background independent" formulation of QG, just because Einstein's equations were universal.   Thinking about this in condensed matter language, this is like saying that there's a "substance independent" theory of condensed matter, just because the NS equations are so universal.  The real clue in Jacobson's work, in a phrase that J.A. Wheeler invented, but clearly did not understand, was that space-time was the IT in "IT from BIT".  The areas of the holoscreens of all causal diamonds determine space-time geometry and these are determined by the von Neumann entropies of quantum density matrices associated with those diamonds.

It follows that one should investigate QG by choosing a background geometry satisfying Einstein's equations with a stress tensor obeying the null energy condition (the second law of thermodynamics), and try to find a quantum system obeying Jacobson's principle in that geometry.  This leads directly to the Hilbert bundle formulation of QG because there is a one to one correspondence between the set of causal diamonds, and the set of choices of nested proper time intervals along time-like geodesics.  The QPR is the natural consistency condition to impose on this system of Hilbert spaces and density matrices.

We next face the question of what to choose for the density matrix of a diamond.  Obviously there can't be a unique choice for this because we know from experiment that there can be many different states inside a causal diamond.  Here we get two clues from quantum field theory and another big boost from the hydrodynamic interpretation of Einstein's equations.  Paradoxically, one of the most important clues is the demonstration by CKN that all of our experiments probe an entropy that scales at most like $(A/G_N)^{3/4}$ (in four dimensions), and cannot account for the entropy used by Jacobson to derive Einstein's equations.  On the other hand, QFT, the {\it theoretical framework} we use to explain our experiments, predicts area law entropy for an {\it empty} diamond, roughly independent of the state in the QFT Hilbert space.  The QFT prediction for the coefficient of area is UV divergent and was conjectured to be "absorbed into Newton's constant" in the Bekenstein-Hawking law.  So we should be looking for a universal prediction for the density matrix of an empty diamond and try to understand the tiny corrections due to excitations localized inside the diamond after we've understood "nothing".

That step was taken by Carlip and Solodukhin in 1998, whose analysis is best understood in an analysis done by the Verlindes, where to leading order in $L/L_P$ a redefined metric is block diagonal between a two dimensional Lorentzian and a $d - 2$ dimensional Riemannian block, with the Riemannian length scales $L \gg L_P^{(d)}$ while those of the Lorentzian block are of order the Planck length in the near horizon limit.  The equations of motion of the large part of the EH action
imply that the two dimensional metric is flat and the Riemannian metric is independent of the two dimensional coordinates.  The two dimensional stress tensor fluctuations of the conformal factor of the Riemannian metric satisfy the Ward identities of a stress tensor of a CFT of large central charge.  That is, they are the hydrodynamic equations of such a CFT.  Carlip and Solodukhin thus postulate that the actual near horizon quantum theory is such a CFT, and show that that postulate reproduces most known black hole entropy formulae.  Zurek and the present author generalized their hypotheses to an arbitrary causal diamond and pointed out that it implied the universal fluctuation formula $\langle (K - \langle K \rangle)^2 \rangle = \langle K \rangle .$

In the present paper we have taken this analysis one step further, using the insight gained from my work with Fischler on HST.  Carlip and Solodukhin do not deal with fluctuations of the unimodular part of the transverse geometry, and Solodukhin comments explicitly that he does not know why it is less important than the conformal factor.  The resolution of this is to imagine that the transverse geometry is encoded in the target space of the CFT. The fact that the finite diamond has finite entropy suggests immediately that the target space fields be fermions\footnote{It's already implicit in the Carlip/Solodukhin work that the CFT lives on an interval and has a cut off spectrum of the Virasoro generator $L_0$.}.
Connes' work on Riemannian geometry and the Dirac equation tells us that there is a natural set of fermion variables lying around, once we remember the spin statistics connection.  This relation was long ago incorporated into HST.   We update it here by turning the spinor fermionic oscillators into $1+1$ dimensional spinor fields.  They're actually Dirac spinors both in the Lorentzian and Riemannian geometries, and so can in a sense be thought of as fields on the full $d$ dimensional space-time.  However, the UV cutoff in the transverse space depends on the proper time in two dimensions in a way that a $d$ dimensional field theorist would find bizarre.  The UV cutoff on the $L_0$ generator is also a bit bizarre.  One takes a central value determined by the classical formula of Carlip and Solodukhin and chooses a width in discrete $L_0$ eigenvalue space such that the Carlip-Solodukhin formula for the entropy is satisfied exactly, for "the smallest causal diamond for which one believes the Carlip-Solodukhin analysis".   That last phrase has unfortunate philosophical implications, to which we will return at the conclusion of these conclusions.  

We are almost done with empty causal diamonds.  To proceed we have to switch our focus from the density matrices of empty diamonds to the time evolution operator between two successive diamonds in a nested covering of a given diamond.  Let us choose a future oriented one \ref{fig:nestedcoversofadiamond} for definiteness.  Here we will restrict attention to geometries that are conformal to maximally symmetric spaces.   For such geometries, the modular Hamiltonian $L_0$ is related\cite{CHM}\cite{JV} to a conformal Killing transformation which leaves the diamond invariant.  More precisely, if we consider a conformal quantum field theory on the same background space-time, then the modular Hamilton of that field theory is proportional to the quantum generator of the action of the conformal Killing vector (CKV) acting on the bifurcation surface of the diamond.  The proportionality constant is fixed by the Virasoro algebra.  The vector field associated with that CKV defines a set of inextensible coordinates inside the diamond, and the flows associated with the vector field are timelike lines, one of which is the geodesic in the diamond.   For these space-times at least, it seems plausible that also in QG we should associate time evolution between two diamonds whose future tips are $\tau$ and $\tau + L_P$ to be
\begin{equation} U(\tau + L_P, \tau) = e^{- i L_0 (\tau + L_P)} . \end{equation} This unitary operator acts in the Hilbert space of the diamond corresponding to the time interval $[- T, \tau + L_P]$ along the geodesic and we expect it to entangle degrees of freedom in the Hilbert space corresponding to $[-T, \tau]$ with the new degrees of freedom that are added in the larger diamond.  It should be tensored with a unitary that acts in the tensor complement of $[-T, \tau + L_P]$ in the full Hilbert space $[-T,T]$\footnote{$T$ is a proper time cutoff inserted to make sure that we are always dealing with finite dimensional Hilbert spaces.  Its value depends on the background space-time.  The limiting cases where the Hilbert space dimension goes to infinity have to be treated with great care.}

We have argued in the previous section that the twin requirements of approximate two dimensional conformal invariance and fast transverse scrambling of information restrict the form of $L_0$ to be that of free fermions plus a limited set of abelian Thirring interactions constructed from quartic fermion operators that would be invariant under transverse area preserving diffeomorphisms if the transverse cutoff were removed.  More work is required to see whether those arguments are indeed correct, but if they are, we have constructed a consistent set of models of quantum gravity for a certain class of space-times.

The most disturbing thing about these models is the fact that we have had to insert an arbitrary restriction to a "smallest causal diamond for which we believe the Carlip/Solodukhin analysis" into our construction.  This is inevitable in an approach that tries to bootstrap quantum theory from hydrodynamics.  By definition, hydrodynamics contains only coarse grained information about an underlying quantum theory.   In principle one should not expect to find the correct description of the microscopic theory applying to the real world, without comparing to microscopic experiments.   This raises the question about models in asymptotically flat or AdS space, where we have a mathematically precise description of asymptotic observables, at least in principle.  In the case of AdS space there seems to be a pretty definitive negative answer to the question of whether we could have an absolutely precise and unambiguous description of the physics inside a causal diamond whose radius was much smaller than the cosmological radius of curvature.  According to the description we have given of that physics in the appendix to this paper, that physics is described by a single node in a tensor network.  But a tensor network is a cutoff quantum field theory and we have known since the work of Wilson that many different cutoff theories converge to the same CFT.   Which one gives the "correct" physics in a small causal diamond, with arbitrary precision?   One can try to impose symmetries to refine the allowed set of cutoff theories, but it seems doubtful that one could come up with a unique definition of time evolution in a sequence of small causal diamonds in terms of unambiguous CFT data.

The same issue arises in asymptotically flat space.  In quantum field theory, we're familiar with the fact that the S-matrix does not determine local correlation functions.  Even in integrable theories in two dimensions one has to supply extra information to determine them, and we can do so only because we know their mathematical definition and properties.  The problem for asymptotically flat quantum gravity is even worse.  For the most part we only know it as a perturbation expansion which is not Borel summable.  There are a few cases where it can be treated by BFSS Matrix Theory, but the large $N$ limit is not under control.  In no case do we have a rigorous understanding of the Hilbert space in which the scattering operator acts.  It seems unlikely that this will lead us to an unambiguous definition of finite time physics in a local region of space-time.

There is a "philosophical" issue at stake here, in which the nasty questions of the interpretation of quantum mechanics raise their heads.  The present author adheres to the modern version of the Copenhagen interpretation of QM.  Certain quantum systems have collective variables whose dynamics takes place on a time scale slower than typical micro-scales.  Hydrodynamic flows are prime examples of such variables.  For collective variables the uncertainties implied by QM are small, down by powers of some ratio of a small length scale to a large one.  More importantly, the violations of Bayes' sum over histories rule for probabilities in the stochastic evolution equations for collective variables are {\it exponentially} small in the ratio of the large length scale over the small one.  This means that we can treat the quantum fluctuations in these quantities like measurement errors or errors caused by random disturbances.    They form the basis for what Bohr and Heisenberg called the Classical World.  In our models, space-time geometry is part of this classical world.

In contemplating the physics that takes place inside a causal diamond, we're imagining a measuring apparatus traveling along some time-like trajectory and "doing experiments".  The meaning of a mathematically precise theory of some set of physical phenomena is that we can account for everything that measuring apparatus can measure, with arbitrary precision.   {\it If the total Hilbert space of the causal diamond is finite dimensional this is simply impossible.} There is some point at which the quantum fluctuations in the measuring device will limit the precision.   Furthermore, in a theory of quantum gravity, when we make the measuring device have more q-bits, in order to endow it with more robust collective observables, it uses up more of the entropy of the diamond, eventually creating a black hole.  So the claim is that there is no such thing as a mathematically precise theory of the physics of a local region of space-time in quantum gravity, and the smaller the region we want to study, the more ill defined the theory becomes.   At a certain point the physics in a "causal diamond" will depend on the nature of the device one uses to probe it.  In language that is probably too classical we can say that the device's gravitational field will distort the local space-time geometry, so that one no longer knows what one was probing.  

For this reason, it seems to me that the phrase "the smallest causal diamond for which we can trust the approximation of Carlip and Solodukhin" represents a fundamental barrier to the construction of local theories of quantum gravity, which can only be removed, if at all, by experimental probes.  That is, the models described in the present paper might one day be tested, and found to apply at large enough scales.  Experiments at smaller scale might then provide guidance about how to construct a more fine grained quantum theory.  Hydrodynamics, as embodied in Einstein's equations and the QPR, can only take us so far.  

I want to end this rather ambitious paper with an outline of how a general theory of quantum gravity looks from the perspective advanced here.
\begin{itemize}
\item Einstein's equations with a smooth stress tensor satisfying the null energy condition are the hydrodynamic equations of the CEP for a general causal diamond.  The c.c. is not included in these equations but is an asymptotic boundary condition which controls the high entropy limit.
\item Generically, causal diamonds have only surface excitations.  Bulk excitations are constrained states.   The empty diamond state has maximal entropy.   In space-times which are asymptotically AdS, these statements are all true on scales below the AdS radius.  The full space-time is a tensor network made up of locally coupled subsystems satisfying these principles.
\item Given a background space-time/hydrodynamic flow, we construct a Hilbert bundle over the space of time-like geodesics on that space-time.  A choice of nested proper time intervals along each geodesic corresponds to a set of nested causal diamonds.  For large enough causal diamonds the modular Hamiltonian of each diamond is the $L_0$ generator of a model of $1 + 1$ dimensional fermion fields, labeled by the eigenspinors of the Dirac operator on the holoscreen of the diamond.  The interactions between these fields is determined by a small number of abelian Thirring couplings, described above.  For space-times conformal to maximally symmetric ones, $L_0$ is the quantum realization of the action of a smooth vector field, restricted to the holographic screen of the diamond.  The flows of that field inside the diamond are time-like and define a set of inextensible coordinates on the interior.  This leads to the natural conjecture that the time evolution operator along those flows, between proper times $\tau$ and $\tau + L_P$ along the geodesic, is just $e^{- i L_0 (\tau + L_P)}$.  The proper times along other time-like trajectories in the coordinate system are shorter than those along the geodesic, so this ansatz properly incorporates the "redshift" between different observers.   A full unitary operator on the Hilbert space over the geodesic is the tensor product of this operator with the $L_0$ of free fermions for those degrees of freedom that are not causally connected to the diamond $[- T, \tau + L_P]$.  For empty diamond states, this ansatz satisfies the QPR.  For constrained states, with localized excitations, the QPR leads, at the level of pictures, to the Feynman-like diagrams we have discussed in the text.  A more quantitative verification of the QPR for localized states is an important unsolved problem.
\item In the limit of asymptotically flat space-time, all of these models are exactly supersymmetric.  That is, their spectrum contains supergravitons and their unitary S-matrix is non-trivial and Poincare invariant. This implies well known results.  The maximum space-time dimension is $11$.  There are a variety of exactly stable BPS branes, and in particular, various compactifications have stable strings with tension much smaller than the Planck scale, so that a systematic perturbation expansion is possible.  Thus, the Hilbert bundle formulation implies conventional string perturbation theory.  

\item The most important features implied by this formalism are those not covered in this paper:  the connection with experiment.  The fluctuation formula $\langle (K - \langle K \rangle)^2 \rangle = \langle K \rangle $, confirms the arguments of Verlinde and Zurek\cite{VZ123etal} that there should be observable quantum gravity fluctuations in interferometer experiments.  The crucial calculation that remains to be done is the power spectral density (PSD), for which only an {\it ad hoc} model\cite{pixellon} exists at present.    I have also argued for years that the connection between the finite horizon of dS space and the breaking of Supersymmetry should lead to insight into the SUSY particle spectrum.  An update on those ideas will appear soon\cite{oldideas2}.  
\item In the appendix I emphasize that the tensor network construction is applicable to quite general CFTs, and that the local picture of AdS space that it provides is not a good guide to the way that bulk locality arises at scales small compared to the AdS radius.  $AdS_d/CFT_{d-1}$ models with EH duals have tensor networks whose nodes are Hilbert spaces with entropy that scales like $R^{d + p}$ where $p \geq 2$.  The reduced density matrices in the node Hilbert spaces, in typical states obtained by acting on the ground state with local lattice operators near the boundary of the (cut off) network, also have large entropy, and Polchinski-Susskind scattering states in the "arena" are constrained, atypical states of the network field theory.  Locality on scales small compared to the AdS radius arises in much the same way as it does in the HST formalism.
\end{itemize}

I want to end this on a personal/philosophical note.  Like most working physicists, I have spent most of my career regarding the problems of the foundations of quantum mechanics in the "shut up and calculate" mode.  We know how to use the theory to make stunningly accurate predictions about observations.  That is all we know on earth and all we need to know.  Controversies that arose during my tenure at UCSC forced me to take a closer look at these issues, as a consequence of which I came away convinced that Bohr and Heisenberg had it essentially right but needed the more modern work on decoherence to justify their somewhat mystical language.  I've explained this in great detail in my textbook, but in brief: certain quantum systems with many DOF and "local" interactions, have a large number of collective variables.  These are operators whose uncertainties scale like inverse powers of a ratio of microscopic and macroscopic length scale, and whose quantum probability distributions satisfy linear stochastic equations up to corrections that are exponentially small in the ratio of the macro to micro length scale.  These variables are the "classical measuring devices" to which the Copenhagen interpretation refers.

The CKN bound on the validity of local field theory in a causal diamond, combined with the apparent inability of black hole dynamics to produce long lived complex collective variables (fast scrambling/incompressible hydrodynamics), suggests that the Copenhagen interpretation is not applicable to the detailed quantum mechanics of gravitational systems.  A local detector can be a complex measuring device, but does not have enough q-bits to store information about the diamond in which it resides.  This implies that mathematically precise descriptions of the physics measureable by that detector are in principle impossible to construct, because they are impossible to test.
For those of us living in an asymptotically de Sitter universe this is somewhat disappointing.

Our mathematically precise theories of AdS space do not help, as discussed in the appendix.  For asymptotically flat space the picture is less clear because we do not have a definition, outside of perturbation theory or the finite N BFSS matrix models of the Hilbert space in which the scattering operator is supposed to be unitary.  Conservatively, one might assume that ambiguities could show up in non-perturbative corrections to the string perturbation series, which is not Borel summable.  As a dramatic example, in \cite{lindil} I proposed modifications of a model of N copies of the minimal Type 0B string theory in $1+1$ dimensions.  These are non-interacting, integrable systems, which have a string perturbation series that is not Borel summable.  Despite the fact that the low energy scattering resembles linear dilaton gravity, there are no linear dilaton black holes in the exact fermionic field theory that gives rise to the perturbation expansion.  The modified models do not effect the perturbation series, but all have meta-stable excitations with the qualitative properties of black holes.  There are a large number of such models (for large $N$), whose detailed predictions are all different.

\section{Appendix The Curious Case of Anti de Sitter Space}

For researchers whose interest in quantum gravity started in String Theory, the focus of interest for 25 years has been the AdS/CFT correspondence.  There are many excellent reasons for this, the principle one being that quantum field theory is our best understood theoretical tool in physics, and this correspondence reduces quantum gravity to the study of a particular limiting case of quantum field theory. I have nothing but admiration for the work in this field and the insights it has brought us, but I believe it has also led to a number of serious misconceptions.

As indicated in the main text, the most serious of these is a completely incorrect notion of the emergence of locality in models of quantum gravity.  This error is both a technical one, and also obscures a fundamental conceptual ambiguity of the notion of local physics in any mathematical theory of quantum gravity.  We will leave the conceptual issue to the end of this appendix and focus on the technical.
The CFT of the AdS/CFT correspondence lives on the boundary $R \times S^{d-2}$ of the universal cover of $AdS_d$, and enjoys the usual locality properties of field theory on that boundary.   One of the earliest and most celebrated properties of the correspondence is Maldacena's {\it scale radius duality}\cite{malda1}.  This implies in particular that if we consider any finite region of $AdS_d$ , below some radius $R*$ in a fixed global coordinate system, we are dealing with some kind of cut off version of the CFT\cite{susswit}.

A very explicit version of the cutoff, which displays the local structure of $AdS_d$ is given by the tensor network (TN) or error correcting code (ECC) models of the AdS/CFT correspondence\cite{swingleetal}.  Commonly referred to as "toy models" these should instead be thought of as sequences of cut off lattice field theories, which converge on the CFT, in a manner that explicitly displays scale/radius duality.  The {\it tensor network renormalization group} (TNRG) of Evenbly and Vidal\cite{swingleetal} makes this even more explicit.  These authors construct unitary embedding maps using a variational principle, which map the lattice model on each shell of the tensor network into that on the next shell.  Remarkably, the eigenvalues of the "Hamiltonians" for small shells, are, in simple soluble CFTs, close to the dimensions of low dimension operators in the CFT on the boundary.

In\cite{tbwfads} Fischler and the present author interpreted the embedding maps of the TNRG as the maps between successive nested causal diamonds along the geodesic that runs through the center of the tensor network.  The shells of the network are the holographic screens of successive nested diamonds.  The TN/ECC picture of CFT thus gives a rather remarkable local picture of the structure of AdS space.  The observant reader will have noticed however that we have never really used the limiting behavior of CFTs that is required in order to make holographic duality work.   This is particularly evident in the work of Evenbly and Vidal, whose numerical studies are done for soluble models with low central charge.  The TN/ECC formalism is a property of QFT, independent of the particular limits which guarantee that a QFT has an "Einstein-Hilbert (EH) dual".  

 There is a universal feature of all known explicit examples of CFTs with an EH dual, namely that one always finds that when the AdS radius $R$ is much larger than all microscopic scales, there are also at least $2$ compact dimensions, whose Kaluza-Klein modes have masses that scale like $R^{-1}$.  Note that it's important here that we say {\it all microscopic scales}.  The free energy of a CFT at temperature $T$ on a $d- 2$ sphere of radius $R$ is
 \begin{equation} F = c T (TR)^{d-2} .\end{equation}  Comparing this to the Bekenstein-Hawking formula for the free energy of a stable black hole in $AdS_d$ we conclude that the AdS radius is large compared to the Planck scale whenever the constant $c$ is large.  A model has an EH dual only if, in addition to large $c$ there is a large gap in dimension between the energy momentum tensor and the typical exponentially growing density of states of a CFT.  In all known examples, the density of states grows too rapidly to be accounted for by a finite number of fields in $AdS_d \times S^1$.   
 
 It has been pointed out in various places, by the present author, as well as by Susskind, that the way to account for this in the tensor network picture is that the nodes of the tensor network must be large quantum systems, and that local physics on scales below the AdS radius, {\it is connected to the behavior of the Hamiltonian within a single node, rather than the local lattice physics of the network}.   There are a number of examples of AdS/CFT with tunable parameters, such that for small values of the parameter a Lagrangian picture of the CFT is valid.   Although this is the opposite limit from the one in which the model has an EH dual, one can examine the "single node" physics in this limit.  It is always a highly non-local "matrix model" (in $1+1$ dimensions this means a permutation orbifold) in which the AdS and compact dimensions appear on a roughly equal footing.  The model exhibits fast scrambling.   On larger scales in the tensor network we find ballistic scrambling on the $d-2$ sphere, obeying a Lieb-Robinson bound, while scrambling in the compact dimensions is still fast.   These properties are mirrored in the hydrodynamics of large black holes in AdS space, in the opposite limit where the system {\it is} well described by an EH dual.  On scales short compared to the AdS radius, hydrodynamics is incompressible and identical to the hydrodynamics of flat space black hole horizons.  There is no hydrodynamic propagation of entropy because information is scrambled at a much faster rate than transverse momentum.  On scales large compared to the AdS radius we find sound modes, which propagate information ballistically.   These dominate the modular fluctuations of large black holes.
 
 These distinctions also show up in the relation between temperature and infall time to the singularity for large and small black holes.  For small black holes, the Schwarzschild radius determines the time scale of infall and also the change in entropy when a particle falls into the black hole.   For a large black hole the particle is falling into one particular node of the tensor network, so the infall time is determined by the timescale of that subsystem, which is the AdS radius.  The temperature on the other hand is a measure of the equilibrium response of the entire network, after the ballistic scrambling process has taken place, and has a very different value.  
 
 A final piece of evidence for the same fact is the difference between the behavior of infinite RT diamonds (AdS Rindler space) and large black holes, both with regard to the coefficient in their modular fluctuation formulae, and the question of the existence of sound modes.  From the bulk point of view this might appear a little puzzling, because the infinite RT diamond has a horizon much larger than the AdS radius, yet it has no sound modes and its modular fluctuations are like those of a flat space diamond.  The resolution of this puzzle is again found by appealing to scale/radius duality and the tensor network construction.  The volume divergence in the entropy of an RT diamond is a  UV divergence in the CFT, and UV divergences have to do with modes of the field theory that are very short wavelength on $S^{d-2}$.  In the tensor network picture, these are modes restricted to a single node, and therefore it is no surprise that they behave like flat space horizon modes.   This example also makes it clear that {\it the sphere at infinity in the CFT 
is not the same as the sphere seen by a local observer in a diamond of radius smaller than the AdS radius.}  The latter sphere, like the compact dimensions, appears in the target space of the field theory, rather than the base space over which the fields are defined.  

 {\it Thus, the emergence of locality at sub AdS radius scales in AdS models with EH duals is not connected with the tensor network and entanglement structure of the boundary CFT.  }   Instead, we claim the picture that emerges is the same as the one we abstracted from the zero c.c. limit of dS space.  The key observation is the "arena" picture of the emergence of the flat space S-matrix from CFT correlators proposed by Polchinski and by Susskind\cite{polchsuss}.  These authors focus on a single causal diamond called the arena whose size is larger than all microscopic scales but smaller than the AdS radius, and construct CFT operators which, when a few of them act on the vacuum, create Witten diagrams that are approximately non-interacting until  all the lines enter into the arena.  The resulting CFT correlators are argued to be directly related to flat space S-matrix elements.   
 
 There are two properties that follow from this picture.   The first assumes the CEP, namely that the arena is a finite dimensional quantum subsystem of the CFT Hilbert space.  By Page's theorem\cite{page} this would seem to imply that for most states of the CFT, the density matrix of the subsystem is maximally uncertain.  It's unclear whether the CEP is true in the exact CFT, but if we identify the arena with the central node in a tensor network approximation to the cut off CFT then it is certainly true and the tensor network shows us that Page's argument is correct for this system.  For a finite tensor network the probability that the density matrix is maximally uncertain is less than one.  It's clear that a lot of care has to be taken when applying Page's theorem to infinite dimensional systems.  
 
 The second property was first pointed out in\cite{tbwfads2} and has to do with the origin of Minkowski amplitudes with large numbers of soft massless particles.  Because of the gap in the AdS energy spectrum for massless particles, one cannot construct Polchinski-Susskind operators that insert an arbitrary number of massless particles into the arena, without creating black holes inside the arena.  Instead, amplitudes with large numbers of Minkowski massless particles arise from AdS correlators of the form
 \begin{equation} \langle PS (1) PS(2) PS(3) PS(4) O(y_1) \ldots O(y_n) \rangle_c , \end{equation} where the $4$ Polchinski-Susskind operators prepare a hard $2\rightarrow 2$ scattering process in the arena, and the other operators create particles whose amplitude to be in the arena (as computed from Witten diagrams) is very small because they are equally probable to be anywhere in AdS space.  We conjecture that it is linear combinations of such amplitudes that converge to S-matrix amplitudes for emission and absorption of arbitrary numbers of soft particles with arbitrarily low energy in the $R\rightarrow\infty$, Minkowski limit.  It's the non-perturbative behavior of those amplitudes for large $n$, which determines whether the resulting states are normalizable in Fock space.  At the moment, it does not appear that the AdS/CFT correspondence sheds additional light on these questions, but the fact that it leads to the same questions as the approach from dS space suggests that the issues raised in the text are real.
 
 We conclude that the issue of local physics on scales smaller than the AdS radius involves a cut-off version of the CFT.   This means that it is not encoded unambiguously in CFT correlators, for many different cut-off systems converge to the same CFT.  In the text we have pointed out a possible deep reason for this ambiguity.  The phrase "physics in a causal diamond of finite area" is shorthand for "a mathematical theory that explains the results of measurements performed by a detector traveling along a time-like trajectory between the tips of the diamond".  However, because the information gathering capacity of a detector is limited by the CKN bound, no mathematically precise theory is testable, so some ambiguity in the definition of local physics is inevitable.   In the real world we could (in principle, though probably not in practice) probe this ambiguity experimentally by showing that the results of experiments depended on the type of detector used.  In the AdS/CFT paradigm the ambiguity seems to be related to the well know insensitivity of continuum field theory to UV details.   This means that the question of precisely what goes on in finite area causal diamonds is not well defined in AdS/CFT.  Different theorists can choose different sequences of cut off models, all of which converge to the same CFT, but differ in their predictions for local physics in finite area diamonds.
 
 \section{Appendix II  Some Remarks about Black Holes in Minkowski Space}
 
 Large black holes in AdS space are conventional thermal ensembles.  Their modular Hamiltonian is just the Hamiltonian of the CFT.  Black holes in Minkowski space have negative specific heat.  One cannot relate their thermal properties to their modular Hamiltonian.  Indeed, the black hole entropy formula would predict a energy level spacing of order $e^{-S}$, while the width of black hole levels (the inverse lifetime) scales like a power of the entropy.  The level spacing of the modular Hamiltonian has nothing to do with the energy of the black hole as measured by an observer at infinity.  We have conjectured instead that it's related to the spacing of levels seen by a near horizon detector that measures the operator $L_0$.  
 
 We should note that the classical properties of black holes beg for an interpretation as some kind of universal equilibrium ensemble.  They are independent of the black hole's formation history and small perturbations of a black hole just lead us to another black hole differing in a small number of macroscopic parameters.  Black holes also admit hydrodynamic flows, which, in the Minkowski case, do not allow for entropy transport.  Our conjecture is a model for this equilibrium state.  
 
 Finally, we note again that negative specific heat gives the simplest resolution of the so called "firewall paradox".  It implies that a large amount of entropy is created when a localized, low entropy, object is dropped onto a large black hole.  The explanation for this is that, in the Hilbert space of the final state black hole, there must have been a large number of frozen q-bits, which must be activated in order for the object to come into equilibrium with the rest of the black hole q-bits.  If the time scale (in proper time of the infalling object) for activation of these q-bits is the Schwarzschild radius, then the infalling object will not begin to feel the effect of being inside the black hole for that amount of time.  Furthermore, the consequence of equilibration is that more and more q-bits with which the object could have interacted if it had been in empty space, will instead be coming into equilibrium.  Using the connection between entropy and area, the effective area "felt" by the object shrinks.

\newpage 
\begin{center}
{\bf Acknowledgments }\\
I would like to thank W. Fischler for collaboration over more than 20 years on material that led to this paper.  Thanks also to B. Fiol for the clue to understanding localized states in quantum gravity.  Special thanks to K. Zurek for showing me the work of the Verlindes and of Solodukhin, which were so important to the formulation of this paper, and for insisting on the universality of the modular fluctuation formula. Thanks also to E. Witten and E. and H. Verlinde for pointing out an error in my description of the Verlindes' work and the way to correct it. This work was supported in part by the U.S. Department of Energy under grant DE-SC0010008
\end{center}

\end{document}